\documentclass[preprint2]{aastex}

\usepackage{psfig}

\newcommand{\msun}{\,\hbox{$M_{\odot}$}}
\newcommand{\kms}{\,\hbox{\hbox{km}\,\hbox{s}$^{-1}$}}
\newcommand{\vrot}{\,\hbox{$V_{\rm rot}$}}
\newcommand{\mbh}{\,\hbox{$M_{\rm BH}$}}
\newcommand{\msigma}{\,\hbox{$M_{\rm BH}-\sigma$}}
\newcommand{\pone}{Paper I}
\newcommand{\degree}{\ensuremath{^\circ}}

\newcommand{\reff}{\mbox{$R_{\rm eff}$}}
\newcommand{\mlr}{\,\hbox{$M_{\odot}$}/$L_{\odot}$}

\shorttitle{Dynamical properties of ULIRGs II}
\shortauthors{Dasyra et al.}

\begin{document}

\title{Dynamical properties of Ultraluminous Infrared Galaxies. II.
Traces of dynamical evolution and end products of local ultraluminous 
mergers\footnote{Based on observations at the European Southern Observatory
(ESO 171.B-044)}}

\author{K. M. Dasyra\altaffilmark{2}, L. J. Tacconi\altaffilmark{2}, 
R. I. Davies\altaffilmark{2}, T. Naab\altaffilmark{3}, 
R. Genzel\altaffilmark{2}, D. Lutz\altaffilmark{2}, E. Sturm\altaffilmark{2},
A. J. Baker\altaffilmark{4,5}, S. Veilleux\altaffilmark{5}, 
D. B. Sanders\altaffilmark{6}, A. Burkert\altaffilmark{3}}







\altaffiltext{2}{Max-Planck-Institut f\"ur extraterrestrische Physik,
Postfach 1312, 85741, Garching, Germany}
\altaffiltext{3}{Universit\"atssternwarte, Scheinerstr. 1, 81679, M\"unchen, 
Germany}
\altaffiltext{4}{Jansky Fellow, National Radio Astronomy Observatory}
\altaffiltext{5}{Department of Astronomy, University of Maryland, College 
Park, MD 20742, USA}
\altaffiltext{6}{Institute for Astronomy, University of Hawaii, 2680 Woodlawn 
Drive, Honolulu, HI 96822, USA}



\begin{abstract}

We present results from our Very Large Telescope large program to study the 
dynamical evolution of local Ultraluminous Infrared Galaxies (ULIRGs) and 
QSOs. This paper is the second in a series presenting the stellar 
kinematics of 54 ULIRGs, derived from high resolution, long-slit $H$- and 
$K$-band spectroscopy. The data presented here, including observations of 17
new targets, are mainly focused on sources that have coalesced into a single 
nucleus. The stellar kinematics, extracted from the CO ro-vibrational 
bandheads in our spectra, indicate that ULIRG remnants are dynamically 
heated systems with a mean dispersion of 161 \kms.
The combination of kinematic, structural, and photometric properties of the 
remnants indicate that they mostly originate from major mergers and 
that they result in the formation of systems supported by random motions,
therefore, elliptical galaxies. The peak of 
the velocity dispersion distribution and the locus of ULIRGs on the 
fundamental plane of early-type  galaxies indicate that the end products of 
ultraluminous mergers are typically moderate-mass ellipticals (of stellar
mass $\sim$10$^{10}$-10$^{11}$\msun).
Converting the host dispersion into black hole mass with the aid of the 
\msigma\ relation yields black hole mass estimates of the order 
10$^7$- 10$^8$ \msun\ and high accretion rates with Eddington efficiencies
often $>0.5$.

\end{abstract}




\section{Introduction}
\label{sec:intro}

Galaxy mergers, which have a frequency increasing with redshift 
(e.g., \citealt{toomre77}; \citealt{kauf93}; \citealt{lefevre}),  
are considered a key mechanism in driving galaxy evolution. In the local 
Universe, the ultraluminous infrared galaxies (ULIRGs) are excellent 
laboratories for studying violent merging events. ULIRGs are mergers of 
gas-rich galaxies observed during strong starburst events; these episodes 
have durations $\lesssim$10$^8$ yrs (e.g. \citealt{canalizo01}; 
\citealt{mihos96}) and are brief compared to the baryonic matter merging 
process ($\sim$10$^9$ yrs; e.g. \citealt{barnes92}; \citealt{hernquist93}). 
The starburst emission, often combined with emission originating from an 
active galactic nucleus (AGN), gives rise to infrared (IR) luminosities 
greater than $10^{12} L_{\odot}$ (e.g., \citealt{sami96}; 
\citealt{lonsdale06} and references therein); these luminosities are 
comparable to the bolometric luminosities of QSOs.

A plethora of studies indicates that ULIRGs transform (gas-rich) spiral 
(S) galaxies into ellipticals (Es) through merger-induced dissipative 
collapse (\citealt{kor92}; \citealt{mihos96}; \citealt{barnes96}). 
The large molecular gas concentrations in the central kpc regions of ULIRGs 
(e.g. \citealt{dowso98}; \citealt{brysco}) have densities comparable to 
stellar densities in ellipticals. The light profiles of ULIRG remnants
often follow a  r$^{1/4}$ law (e.g., \citealt{wright}; \citealt{stanford}; 
\citealt{james}; \citealt{sco00}; \citealt{rothberg04}). \cite{kim02} and 
\cite{veilleux02} established the solidity of this result by proving that 
the majority (73\%) of all single-nucleus
ULIRGs in a sample of 118 sources with 60 \micron\ 
flux greater than 1 Jy (hereafter the 1 Jy sample; \citealt{kim98}) have 
elliptical-like light profiles. \cite{genzel01} and \cite{tacconi02} have made 
high-resolution near-infrared (NIR) spectroscopic measurements of the stellar 
dynamics of small samples consisting mostly of fully-merged ULIRGs.  They 
conclude that ULIRGs resemble intermediate mass ellipticals/lenticulars with 
moderate rotation, in terms of their velocity dispersion distribution, their 
location in the fundamental plane (FP; e.g., \citealt{djoda}; 
\citealt{dressler}) and their distribution of the  rotation/velocity 
dispersion ratio. Together, these results suggest that ULIRGs form moderate 
mass ellipticals (of stellar mass $\sim$10$^{11}$ M$_{\odot}$).

One way to investigate the physical details and the evolution of ULIRGs
is to determine the kinematic and structural properties of the
merging galaxies in different merger phases. We therefore 
conducted a European Southern Observatory (ESO) large 
program\footnote{(PI Tacconi)} that traces the host dynamics of a 
large sample of ULIRGs spanning wide ranges of merger phase and infrared 
luminosity through NIR spectroscopy. Our study extends the previous work of 
\cite{genzel01} and \cite{tacconi02}. The enlarged sample comprises ULIRGs
at wider ranges of merger phase and luminosity, and higher redshifts than 
those of \cite{genzel01} and
\cite{tacconi02}. The new observations of 38 sources increase the 
original sample by more than a factor of 3, enhancing our statistics and 
enabling us to study the properties of galaxy mergers as a function of time.
In Dasyra et al. (2006, hereafter \pone) we analyzed those ULIRGs that are 
in a merger phase later than the first encounter but prior to nuclear 
coalescence, and hence, show more than one nucleus in the NIR acquisition 
images. In this paper we present results  mainly from those ULIRGs 
that have coalesced and show a single nucleus in our images, the so-called 
remnants. We compare the stellar kinematic properties of binary ULIRGs and 
ULIRG remnants to look for traces of dynamical evolution in 
ultraluminous mergers. 

Some of the ULIRGs presented in this study may in fact be binary sources 
very close to coalescence that have small projected nuclear separations;
such sources cannot always be resolved or kinematically disentangled due to 
instrumental angular resolution constraints. At redshifts typical for the 
sources in our sample, the angular resolution achieved implies that any 
unresolved systems will have nuclear separations smaller than 1.5 kpc. 
Merger simulations (e.g. \citealt{mihos99}; \citealt{mihos00}) 
have shown that by the time the individual nuclei are separated by $\lesssim$ 
1 kpc, the low moments of the stellar kinematics (rotational velocity and
velocity dispersion) have almost reached their relaxation values. Therefore, 
the dynamical properties of all the sources we classify as remnants are 
representative of those at dynamical equilibrium. 

This paper is arranged as follows.
We briefly summarize the observations and data reduction methods and 
present the pre- and post- coalescence ULIRG samples in \S~\ref{sec:obs}.
After studying whether the kinematic properties of the merging galaxies evolve 
during the ultraluminous merger phases in \S~\ref{sec:host}, we investigate  
the origins and the potential end products of ultraluminous mergers in 
\S~\ref{sec:origin} and \S~\ref{sec:end} respectively. We then focus on
the black holes hosted by ULIRGs: an analysis of the evolution of 
the \mbh - $\sigma$ relation during the merger is followed by a discussion 
of the nuclear activity implied by our data in \S~\ref{sec:black}. Finally 
a summary is presented in \S~\ref{sec:conc}.


\section{Observations And Data Reduction}
\label{sec:obs}

\subsection{Data Acquisition And Analysis}

We combine the sources in our Very Large Telescope (VLT) program 
with the sources presented in \cite{genzel01} and \cite{tacconi02} to
compile a sample that comprises 54 ULIRGs and 12 QSOs. 
The stellar kinematics of the QSOs will be presented in a forthcoming paper 
(Dasyra et al. 2006, in preparation). In this paper, we perform a study 
of the stellar kinematics in post-coalescence ultraluminous mergers (ULIRG 
remnants), after presenting new data for 17 ULIRGs and summarizing 
pre-existing data for 13 mainly post-coalescence  sources. Of the 54 
ULIRGs observed in total, 30 are confirmed to have two nuclei\footnote{20 
were first presented in \pone\, and 10 are presented or summarized here; see 
the Appendix}, 
and 1 (IRAS 00199-7426: \citealt{duc97}; \pone) may be a multiple-interaction 
system. A detailed description of the criteria set to select these 54 sources 
from the 1 Jy catalog is given in \pone.

Our new high-resolution, long-slit spectroscopic data were obtained 
using the ISAAC spectrometer (\citealt{moor98}) mounted on the Antu 
telescope unit of the VLT. The observations were made in the $H$ and $K$
bands, with an instrument resolution of $\lambda / \delta \lambda$=5100 
and $\lambda / \delta \lambda$=4400 respectively.
The slit length was 120\arcsec\ and the slit width 0\farcs6; the 
detector scale was 0\farcs146 per pixel. The total exposure times 
and the slit position angles for each source are presented in 
Table~\ref{tab:list}. If the position angle of the major axis of rotation 
could not be identified for the remnants (e.g. from the elongation of the 
stellar disk), the slits were typically placed at 0\degree\ and 90\degree. 
The redshift range of all sources in our sample is $0.018<z<0.268$ (see 
Table~\ref{tab:list}). No sources are observed for the redshift sub-range 
$0.163<z<0.199$, since the CO bandheads are then shifted into wavebands of
high atmospheric absorption.

As in \pone, we derive the structural parameters of the sources that do 
not have high-resolution NIR imaging (e.g. Hubble Space Telescope NICMOS) data
by fitting ellipses to our 10-second-long acquisition images.
The acquisition images have also been obtained with the ISAAC detector, in its 
$H-$band imaging mode, which corresponds to a scale of 0\farcs147 per pixel. 
We fit the ellipses with the aid of the SExtractor package (\citealt{bertin})
and we present the half-light radius $R_{\rm eff}$ \footnote{As measured
from the ellipse that encloses half the total counts.}, the ellipticity, and 
the angle $\phi_{\alpha}$ between the major axis of rotation and the position 
angle of the first slit in Table~\ref{tab:structure} for each source. To 
convert all angular distances into linear sizes we use a 
$H_0$=70~km~s$^{-1}$~Mpc$^{-1}$, $\Omega_{m}$=0.3, $\Omega_{\rm total}$=1 
cosmology. For some of the sources in our sample, a measurement of the 
$K$-band effective radius is available in the literature (see 
Table~\ref{tab:structure}). In these cases, we use the average NIR \reff\ 
value in all computations and diagrams that follow. The imaging analyses of 
\cite{surace99} 
and \cite{sco00} have shown that the difference between the measurement of 
\reff\ in ULIRGs in the $H$ and in the $K$ band is small and non-systematic. 
Therefore, averaging these NIR \reff\ values yields a measurement of higher 
accuracy.

\begin{figure*}
\centering
\includegraphics[height=20cm,width=16cm]{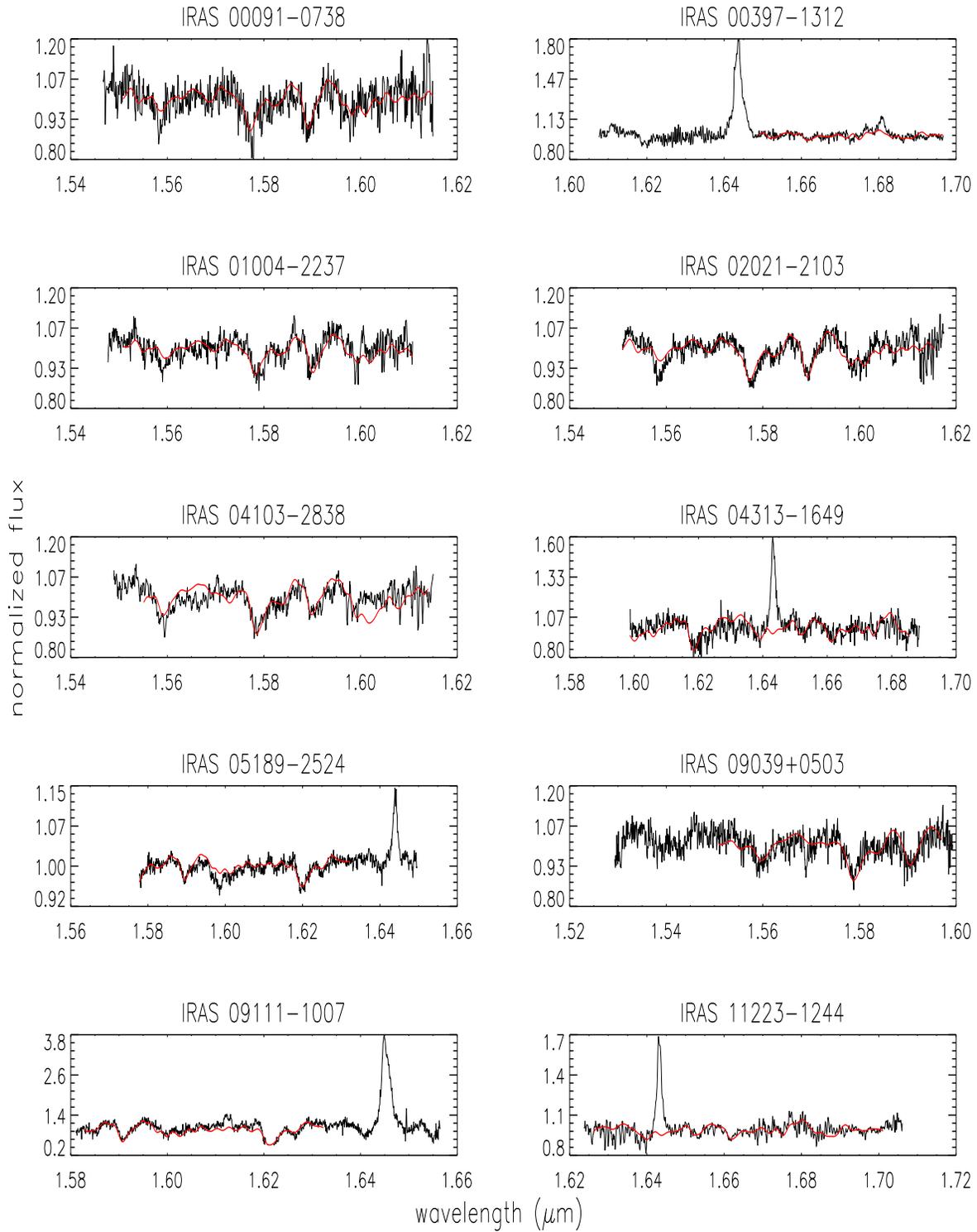}
\caption{The $H$-band spectra of the ULIRG remnants from this study. The 
stellar template, convolved with Gaussians that represent the LOS broadening
function of the sources, is overplotted as a solid line. All spectra are 
shifted to rest frame. 
\label{fig:spectra} }
\end{figure*}

The extraction of the stellar central velocity dispersion $\sigma$ and 
rotational velocity $V_{\rm rot}$ from the spectra follows the method 
presented in \pone. It is performed using the Fourier quotient technique 
(\citealt{bender}); this method provides the intrinsic line-of-sight (LOS) 
velocity profile along any given aperture. To this we fit a high-order 
Gaussian (linear combination of a Gaussian and a second order polynomial)
to determine the average LOS radial velocity and velocity dispersion.  
The fit is performed to each bandhead individually and the errors are equal 
to the standard deviation of all measurements performed. The central 
aperture spectrum of each source, combined over the two slits and 
shifted to restframe, is displayed in Fig.~\ref{fig:spectra}. The stellar 
template that fits best the post-coalescence ULIRG spectra is an M0III giant 
(HD 25472; presented in \citealt{genzel01}). In general, template mismatch
affects the accuracy to which the stellar velocities are measured; we 
found that the difference in the velocity dispersion between the 
(best-fitting) M0III giant and the M1I supergiant (HD 99817;  
\citealt{genzel01}) is $\sim$15 \kms. The M0III template is 
overplotted with a dashed line in Fig.~\ref{fig:spectra}, after being 
convolved with the Gaussian that best simulates the LOS broadening function.

In all of this paper's analyses we deem aperture effects to be 
negligible for the measurement of the stellar velocity dispersion.
According to the merger simulations of \cite{beba00}, the velocity 
dispersion in (relaxed) remnants may vary at most by 10\% for the apertures 
considered here (up to $\sim$2\reff) which are selected to maximize the 
signal-to-noise ratio. Resolution effects further reduce differences 
in the velocity dispersion measurement between various apertures.

The LOS rotational velocity, $V_{\rm rot}{\rm (LOS)}$, is measured from 
apertures that exclude the center of the galaxy. The center and the 
annular width of the apertures used are tabulated in 
Table~\ref{tab:velocities}. For each slit, the observed rotational velocity 
$V_{\rm rot}(obs)$ can be derived from the LOS rotational velocity by 
correcting for the angular deviation of the slit from the major axis 
of rotation as 
\begin{equation}
\mbox{$V_{\rm rot}(obs)=V_{\rm rot}{\rm [LOS]}/ cos[\phi_{\alpha}]$.}
\end{equation}
When rotation is observed along both slits, i.e. when none of the
slits is very close to being perpendicular to the major axis of rotation, 
we average the results of the two slits. The value of the 
observed rotational velocity is given in Table~\ref{tab:velocities} 
and is related to the actual rotational velocity $V_{\rm rot}$ by
\begin{equation}
\mbox{$V_{\rm rot}{\rm (obs)} = V_{\rm rot} sin(i)$}. 
\end{equation}
In this paper
we do not compute the inclination $i$ from the ellipticity of each source (as 
we did in \pone); this conversion is very uncertain for the ULIRG remnants 
since their stellar disks are dynamically hot (due to the advanced phase of 
the merger). Instead, we statistically determine the mean inclination that
needs to be applied to the sample. For this purpose, we use the weighted 
mean value of $sin(i)$ observed for disk-like galaxies on the sky. 
Since the probability $p[i]$ of finding a galaxy at an inclination i (in 
the range [0\degree,90\degree]) scales with $sin(i)$ (\citealt{collin}), 
the mean value of $p[i]sin(i)$ is $\pi /4$. The stellar kinematic results 
(central velocity dispersion and rotational velocity) can be found in 
Table~\ref{tab:velocities}.


\subsection{Pre- and Post-coalescence Sample Classification}
\label{sec:selection}
Prior to performing statistics on the kinematic parameters of the pre- and 
post- coalescence ULIRG samples, we descibe the criterion we use to 
compile the two samples; it consists of a nuclear separation threshold
beyond which binary sources can be considered relaxed. According to 
simulations, this nuclear separation should be roughly 1 kpc 
(\citealt{mihos00}). The actual value of the threshold we set depends 
on the resolution of the images we use to search for the presence of
secondary components. 

We compute the resolution of our images, which depends on the detector scale 
and the seeing, by averaging the FWHM of bright stars in each field. 
Our mean resolution corresponds to a FWHM of 4.3 pixels and enables 
us to resolve individual sources separated by 1 kpc up to $z$=0.080. If we 
set the nuclear separation threshold to 1.5 kpc, we can distinguish 
individual sources up to $z$=0.118. The corresponding numbers of sources
that are verified to satisfy the coalescence criterion are 10 and 23, 
respectively. We adopt a nuclear separation cutoff of 1.5 kpc. For this 
nuclear separation, the low velocity moments are still close to their 
relaxation values according to the models of \cite{mihos00} and \cite{naab06}.

\begin{figure*}
\centering
\includegraphics[width=14cm]{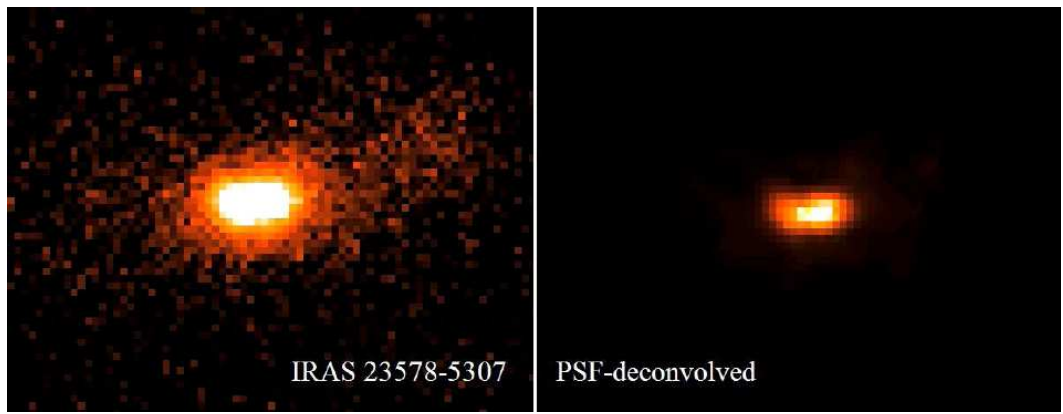}
\caption{The ISAAC $H$-band raw ({\it left panel}) and PSF-deconvolved
({\it right panel}) acquisition image of IRAS 23578-5307.
\label{fig:onesource} }
\end{figure*}

Of the 7 sources at $z>0.118$, 5 have been observed with the NICMOS camera 
onboard the Hubble Space Telescope by \cite{veilleux06} and \cite{sco00}.
According to these authors, none of IRAS 00397-1312, IRAS 01572+0009, 
IRAS 04313-1649, IRAS 09039+0503, and IRAS 14070+0525 has a secondary object 
at distances $<$1.5 kpc that can be unambiguously characterized as a nucleus. 
For one of the remaining 2 sources, IRAS 23578-5307, the large elongation 
and the tidal tails indicate the possible presence of two components 
(see Fig.~\ref{fig:onesource}; left panel). To investigate
this possibility, we deconvolve our acquisition image with its 
point-spread-function (PSF) which has a FWHM of 4.1 pixels. For this 
task, we use the LUCY algorithm of IRAF (\citealt{lucy}; \citealt{ricky}). 
The deconvolution increases the resolution of the image by a factor of 
$\sim$2. The resulting image (see 
Fig.~\ref{fig:onesource}; right panel) confirms the presence of a second 
nucleus and constrains its projected separation to be 1.4 $\pm$ 0.4
kpc. Unlike that of IRAS~23578-5307, the redshift of IRAS~11223-1244 is too 
high to enable us to derive any conclusions using image deconvolution 
techniques. IRAS~11223-1244 is therefore excluded from all statistics that 
follow. The same applies to the possibly multiple merger IRAS~00199-7426. 

A summary of the classification of all sources in our sample can be found
in Table~\ref{tab:list}. We have a total sample of 29 ULIRG remnants,
which consists of 21 single-nucleus or unresolved sources, and 8
confirmed close binaries that have nuclear separations less than or equal 
to the selected threshold. These sources namely are IRAS~00091-0738, 
IRAS~09111-1007, IRAS 14378-3651, IRAS~15250+3609, IRAS~23578-5307, 
Mrk~273, Arp 220 and NGC 6240 (see the Appendix). The pre-coalescence 
binary sample comprises 23 pairs of sources; of those, 40 
individual components have velocity dispersion measurements.


\section{Traces of Evolution In The Stellar Kinematics}
\label{sec:host}

\begin{figure*}
\centering
\includegraphics[width=6cm,angle=270]{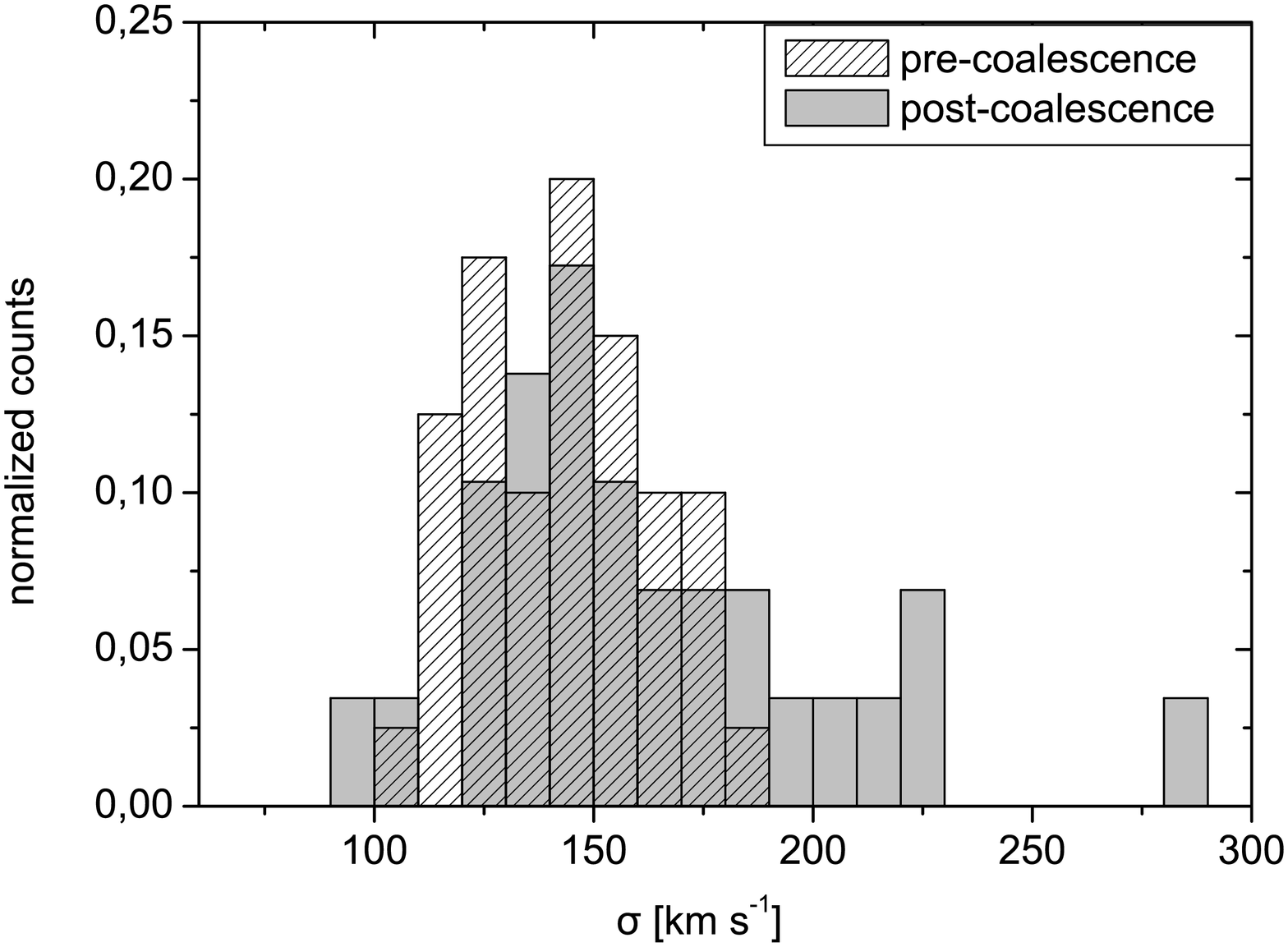}
\includegraphics[width=6.23cm,angle=270]{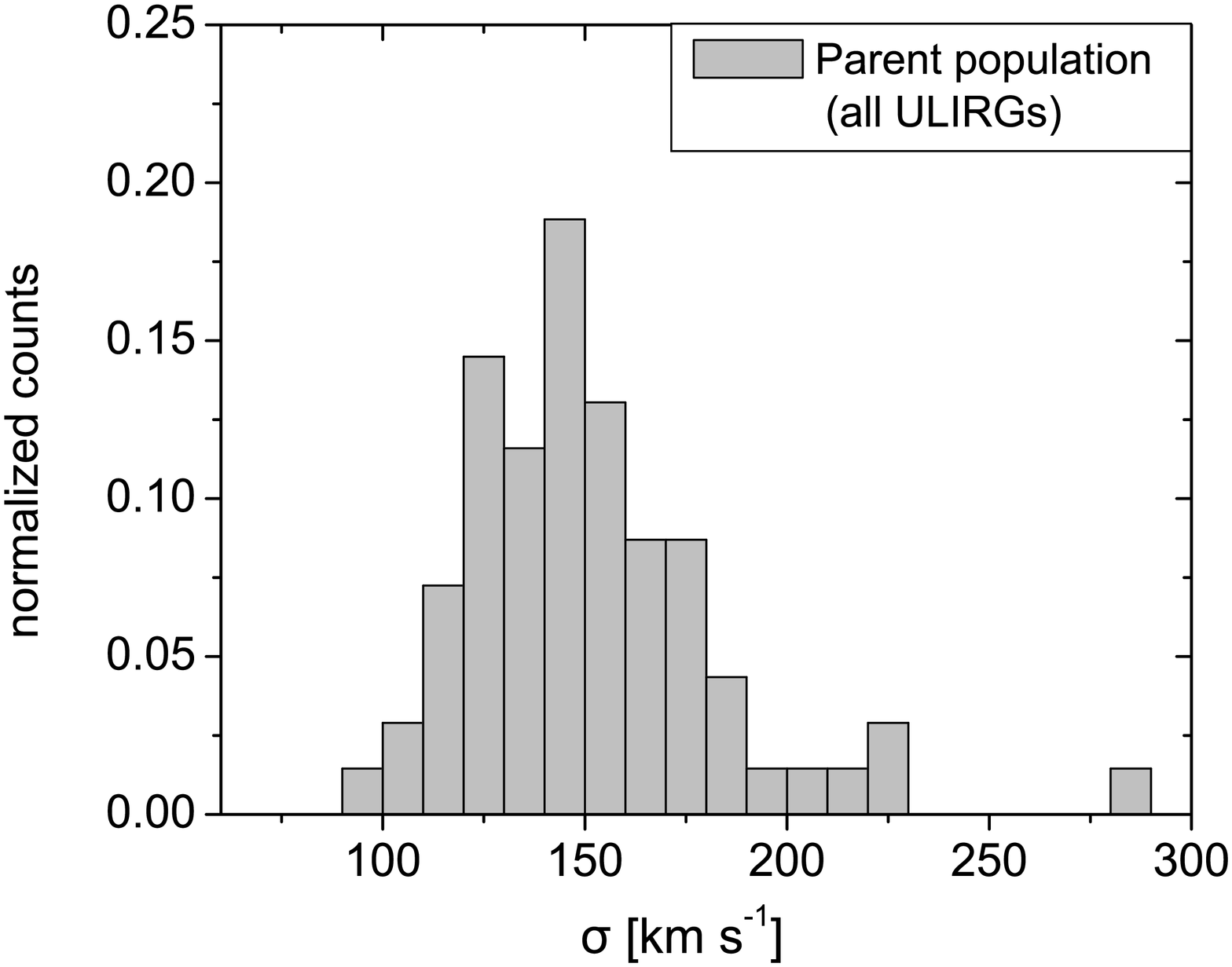}
\caption{
{\it Left panel:} The distributions of stellar dispersions in ULIRGs before 
and after nuclear coalescence.
{\it Right panel:} The ULIRG parent-population distribution used in our 
Monte Carlo simulations. 
\label{fig:distrib} }
\end{figure*}

Taking into account the classification scheme of \S~\ref{sec:selection}, we 
find that the $\sigma$ distribution of the 29 ULIRG remnants has a mean of 
161 \kms\ and a median of 150 \kms. The uncertainty in this mean, which is 
known as its standard error, is 8 \kms\ since the standard deviation equals 
42 \kms. For the 40 pre-coalescence ULIRG nuclei, we find that the mean is 
142 \kms, with a standard error of 3 and a standard deviation of 21 \kms, 
and the median is 145 \kms. The pre- and 
post- coalescence dispersion distributions are shown in Fig.~\ref{fig:distrib} 
(left panel). While the difference in the mean of the two distributions 
is small and the modes are the same, the remnant velocity dispersion 
distribution has a larger variance and a tail at the high-$\sigma$ end, the 
statistical significance of which needs to be quantified. 

To investigate whether the two distributions are independent, we begin
by using the Kolmogorov-Smirnov (KS) test, which makes no assumptions about
the shape of the distributions under examination. We find that the maximum 
deviation $D$ between the cumulative distribution function of the two ULIRG 
populations is 0.285, which for the number of sources we observed, corresponds 
to a probability of 89.4\% that the populations are independent. This 
probability is below the widely accepted significance levels. However, a 
well-known disadvantage of the KS test is that it is sensitive to small 
number statistics. To address whether a better handling of small number 
statistics may alter our conclusions, we created Monte Carlo simulations.

The Monte Carlo code begins with the assumption that the two distributions
are drawn from the same parent population. We simulated the parent 
distribution by spline interpolating between the observed $\sigma$ 
distribution of the combined pre- and post- coalescence ULIRGs (see 
Fig.~\ref{fig:distrib}, right panel). From the parent distribution,
which consisted of 10$^6$ points, we selected random points to generate 
two artificial progenitor/remnant subsamples. Each generated subsample 
had a number of elements equal to that of the corresponding real population. 
We repeated this procedure for 10000 iterations. In each iteration, we 
computed the difference in the mean of the two generated samples and the 
difference in the variance between each real population and the respective 
generated sample. At the end of all iterations, we counted how many 
times the difference in the mean of the two generated samples was 
equal to or greater than that measured, 19 \kms. We found that it 
corresponded to a probability of 1.2\%. Similarly, the probability $p$ 
that the variances of both generated samples ($\sigma^2_{\rm gen,1}$ and 
$\sigma^2_{\rm gen,2}$ respectively) are greater than those measured (or 
\mbox{$p[\sigma_{\rm gen,1} > 21\kms]*p[\sigma_{\rm gen,2} > 42\kms]$}) 
is 5.0\%. In other words, the different means imply a 98.8\% 
probability that the two distributions are independent, while the different 
variances yield a 95.0\% probability of independence. Since these 
probabilities correspond to $\sim$2.5 and $\sim$2.0 sigma 
respectively, the difference in the distributions is at or above the 
widely used significance levels. This result favors the hypothesis that the 
apparent kinematic evolution is real rather than an artifact of limited-number 
statistics. 

Despite the large range of confidence levels indicated by the 
statistical tests, their conclusion can be summarized as follows: it is 
possible, although uncertain, that the observed increase in sigma reflects the
dynamical heating of the merging galaxies. The importance of this result
lies in that observations are able to constrain the dynamical heating 
that models predict for the ultraluminous phases of gas-rich mergers. 
Theory (\citealt{mihos96}; \citealt{mihos99}; \citealt{springel05}) and 
observations (\citealt{bothun}; \citealt{murphy}; \citealt{veilleux02}) 
show that strong starbursts (and associated ULIRG phases) in gas-rich mergers 
typically occur between first encounter and nuclear coalescence, altough 
in individual cases, their actual occurence may significantly vary in terms of 
time. On a statistical basis, a binary ULIRG sample is expected to be roughly
mid-way from first encounter to coalescence; a ULIRG remnant sample is probably
close to coalescence. For these merger phases, the theoretically predicted 
increase of the stellar velocity dispersion is small (\citealt{naab06}); 
therefore observationally tracing it is very important.
However, prior to comparing observations with models, further data are 
needed to quantify at 3 Gaussian-sigma levels the difference in 
$\sigma$ before and after nuclear coalescence.

Another possibility that needs to be investigated prior to comparing the
evolution in the stellar kinematics between observations and models is
whether the double- and single- nucleus ULIRGs may originate from mergers
of galaxies of somewhat different mass contents: the ULIRGs that are prior 
to nuclear coalescence may be more gas-rich (and possibly more massive) 
than the remnants, allowing them to have an ultraluminous phase of comparable 
IR output at earlier merger phases (\citealt{mihos96}). In this case, the 
observed dynamical heating would be less than what would be measured if we 
were able to observe a specific galaxy pair from the beginning until the end 
of the merging process.  To investigate whether the starburst activity 
is triggered under similar conditions for pre- and post- coalescence ULIRGs, 
we need to quantify and compare the gas-mass content of both samples. 
Molecular gas mass measurements obtained by \cite{gao99} in local Luminous 
Infrared Galaxies (LIRGs; sources of $10^{11} L_{\odot}< L_{\rm IR} 
<10^{12} L_{\odot}$) and ULIRGs indicate that a correlation between nuclear 
separation and gas fraction is observed in LIRGs but not in ULIRGs.  
To properly address the question, molecular-gas-content observations 
need to be performed on ULIRG samples with sizes comparable to those of 
the dynamical studies.

The mean value of the rotational velocity is 62 (and the standard 
deviation is 59) \kms\ for the ULIRG remnants, increasing to 79 \kms\ 
when inclination effects are statistically accounted for. The observed 
stellar rotational velocity of each individual source and its ratio over the 
velocity dispersion is presented in Table~\ref{tab:velocities}. The mean
$V_{\rm rot}{\rm (obs)}/\sigma$ ratio of the remnants is 0.36, increasing 
to $V_{\rm rot}/\sigma$=0.46 when we apply the statistical inclination 
correction to the rotational velocity. The individual progenitors of the 
binary systems have a mean rotational velocity of 59 (with a standard 
deviation of 38) \kms\ or 105 (with a standard deviation of 96) \kms\ when 
the inclination effects are corrected from the ellipticity of each 
progenitor. The corresponding observed and 
inclination-corrected $V_{\rm rot}/\sigma$ ratios of the progenitors are 
0.42 and 0.76. The somewhat lower $V_{\rm rot}/\sigma$ ratio of the remnants 
can only be attributed to their higher (than the progenitors) value of 
$\sigma$, since the difference between the pre- and post- coalescence 
rotational velocity is insignificant.
In these calculations we have not attempted to correct the central velocity 
dispersion for inclination effects.

\section{Origin Of The ULIRG Remnants}
\label{sec:origin}

We infer the initial conditions of the mergers that lead to
ultraluminous IR activity by comparing the 
kinematic properties of ULIRG remnants with those predicted by simulations 
in the literature. According to various authors (e.g. \citealt{beba00};
\citealt{naab03}) the remnant \mbox{$V_{\rm rot}{\rm (obs)}/\sigma$} 
ratio depends upon the initial mass ratio of the merging galaxies.
In the gas-free, N-body simulations of binary mergers performed by
\cite{naab03} for several mass ratios and orientations, the major mergers 
produced slowly rotating remnants. Those authors suggested that the 
$V_{\rm rot}{\rm (obs)}/\sigma$ ratio is $\sim 0.2$ for 1:1 and $\sim 0.4$ 
for 2:1 merger remnants, while it reaches higher values (0.8) when the 
remnants originate from minor (4:1) mergers. Our remnants agree best
with a 1:1 and 2:1 merger origin, also agreeing with the directly measured
progenitor mass ratios of \pone\ and confirming that ULIRGs are representative
of the most violent local mergers. The results are similar when comparing to
those of \cite{naab06}, who have presented similar contours for gas-rich, 
dissipative simulations of 1:1 and 3:1 mass ratios.

\begin{figure*}
\centering
\includegraphics[width=14cm]{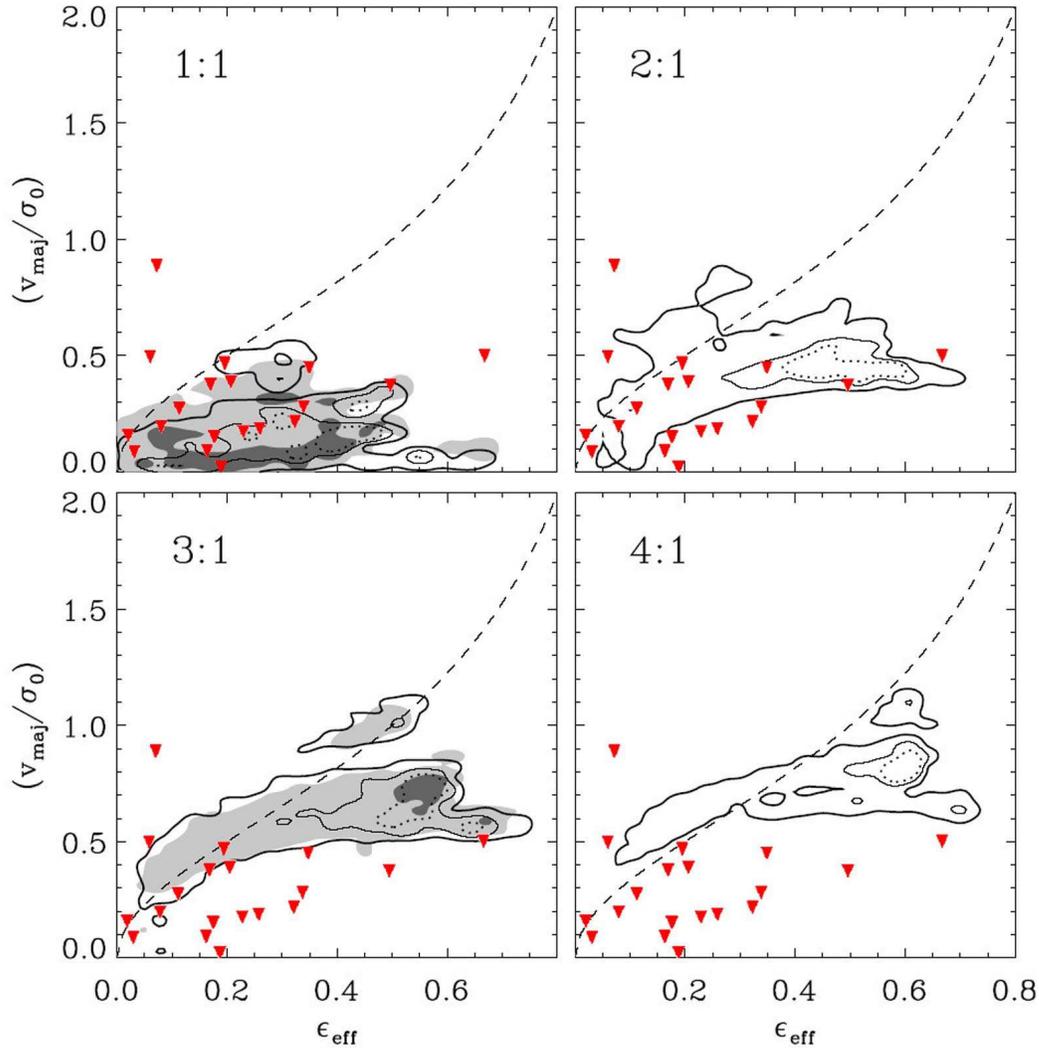}
\caption{Ratio of observed rotational over dispersion velocity versus 
ellipticity for merger remnants (from \citealt{naab03} and \citealt{naab06}). 
Each panel 
corresponds to mergers of different progenitor mass ratios (cases from 1:1 
to 4:1 are studied). The bold line, solid line, and dotted contours correspond
respectively to the 90\%,70\%, and 50\% probability of finding a collisionless 
merger remnant in the enclosed region. The dark- and light-gray shaded areas
indicate the region in the diagram where gas-rich merger remnants are 
expected to be found at 90\% and 50\% probability levels. The dashed line 
shows the theoretical values for an oblate isotropic rotator. ULIRG remnants
are shown as triangles.
\label{fig:ell}}
\end{figure*}

\cite{naab03}, \cite{burkert05}, and \cite{naab06} have also shown a connection
between the $V_{\rm rot}{\rm (obs)}/\sigma$ ratio and the remnant ellipticity 
at the effective radius for several progenitor mass ratios. We overplot our 
results with those of the merger simulations in Fig.~\ref{fig:ell}, where our 
ULIRGs are given in triangles. Each panel corresponds to mergers of different 
progenitor mass ratios (denoted at the upper-left corner of each panel). In 
bold, solid, and dotted contours we show the 90\%,70\%, and 50\% probability 
of finding a merger remnant of each category in the enclosed region. These 
contours correspond to models that do not include gas (\citealt{naab03}). For 
the dissipative merger remnants of \cite{naab06}, the 90 \% and 50\% 
probabilities are shown as dark- and light-gray shaded areas respectively. 
A comparison of the predicted with the measured remnant dynamics indicates 
that ULIRGs are more likely to be formed by major mergers (see \pone).
We also find that ULIRGs have a narrower mass-ratio origin than the merger
remnants in \cite{rothbergb}, who have carried out a similar analysis on a 
sample of sources that are selected visually to have perturbed morphologies 
(see also \S~\ref{sec:disc}).

The addition of gas to the simulations of \cite{naab03} did not 
significantly influence the position of remnants on the 
$V_{\rm rot}{\rm (obs)}/\sigma$ vs. $\epsilon$ diagram (\citealt{burkert05};
\citealt{naab06}). However, the addition of star formation may 
result in an increase of the theoretically predicted $V_{\rm rot}{\rm (obs)}
/\sigma$ ratio and ellipticity, since new stars will be created upon a 
dynamically-cold disk, formed by gas that is not consumed during the merger 
(\citealt{barnes02}; \citealt{springel00}). 
This possible increase would shift the probability contours to higher 
values on both axes further supporting our findings that ULIRGs are produced 
mainly by mergers of galaxies of comparable masses.  On the other hand, the
position of ULIRGs on this diagram could be shifted upwards along the
\vrot /$\sigma$ axis due to aperture correction effects. The \vrot\ values 
used are not necessarily measured from apertures where the rotation curve has 
reached its flat part (\citealt{vanalbada}). However, the facts that \vrot\
has been measured from apertures between 0.5 and 2 \reff\ and that the
ellipticity at the effective radius does not depend upon the apertures 
used to measure \vrot\ are good indications that such  corrections will 
probably not be sufficiently large to alter our conclusions.

\section{End Products Of ULIRGs}
\label{sec:end}

\subsection{The Masses Of Ultraluminous Merger Remnants}
\label{subsec:42}

A direct way to investigate the nature of the end products of  ultraluminous 
mergers is to calculate the mass $m$ of their remnants. According
to \cite{bender92} the bulge mass is related to the stellar kinematics as 
\begin{equation}
\mbox{$m = c_2 \sigma_{100}^2 R_{\rm eff}$,}
\end{equation}
where $\sigma_{100}$ is the projected central velocity dispersion in units of 
100 \kms, $R_{{\rm eff}}$ is in kpc, and $m$ is in 10$^{10}$ \msun. 
The geometrical factor $c_2$ depends on the distribution of matter in the 
galaxy. Following \cite{tacconi02}, we adopt $c_2=1.4$, as appropriate 
for a constant $m$/L King model whose tidal-to-core radius ratio
is 50, midway between those of dwarf and giant ellipticals. By combining the 
above, the total dynamical mass is computed from 
\begin{equation}
\mbox{$m=4.7\times 10^5 (3 \sigma^2 + \vrot^2) R_{{\rm eff}}$,} 
\end{equation}
where $\sigma$ and \vrot\ are now in units of \kms\, \reff\ is in kpc, 
and $m$ is in \msun. 

We use the mean values of the stellar dispersion and inclination-corrected 
rotational velocity from \S~\ref{sec:host}. For the objects that do not have
a \vrot\ measurement (see Table~\ref{tab:velocities}), we use their dispersion 
in combination with the mean inclination-corrected \vrot/$\sigma$ ratio (0.46) 
to infer it. We find that the mean dynamical mass of the remnants is 
8.91$\times 10^{10}$ \msun (in good agreement with \citealt{tacconi02}), 
suggesting that ultraluminous activity mainly originates from mergers of 
sub-$m_*$ galaxies, for $m_*=1.4\times10^{11}$ \msun\ (\citealt{genzel01} and 
references therein). For different values of $m_*$ (e.g that of 
\citealt{bell03} adapted to our cosmology, $8.3\times10^{10}$ \msun) the 
sub-$m_*$ characterization of ULIRGs may change to $\sim m_*$.
Still, in the local Universe, the disk progenitors of ULIRGs
do not need to be as massive as e.g. the Milky Way. 

ULIRG masses derived from H${\alpha}$ emission-line dispersions are
also sub-m$_*$ (\citealt{colina}). Tracing the ionized gas by the [FeII] 
emission lines that appear in our spectra, we find that while the mean 
dispersion of the gas is similar to that of the stars, in individual cases 
gas and stellar dispersions may significantly deviate due to gas outflows. Gas 
kinematics in these ULIRGs will be presented in a forthcoming paper (Tacconi 
et al. 2006, in preparation).

The result that ULIRGs are sub-$m_*$ to $m_*$ galaxies does not 
contradict findings from imaging studies that ULIRGs 
have NIR luminosities greater than $L_*$ (e.g. \citealt{sanders00}; 
\citealt{colina01}; \citealt{veilleux06}). ULIRGs can simultaneously 
be sub-$m_*$ and (1-2)$\times L_*$ (in the $H$-band) mergers 
since they are selected at the peak of their starburst, at which time 
their luminosity-to-mass ratio rises (see also \citealt{tacconi02}; Rupke 
et al. 2002; 2005a; 2005b).

\subsection{ULIRGs And The Fundamental Plane Of Early-type Galaxies}
\label{subsec:fp}

The $V_{\rm rot}/\sigma$ ratios of the ULIRG remnants indicate that mergers
of ultraluminous infrared output lead to random-motion-dominated systems with a
non-negligible rotational component. To investigate what type of ellipticals
ultraluminous IR mergers form, \cite{genzel01} and \cite{tacconi02} placed
our initial sample of ULIRGs on the fundamental plane of early-type
galaxies (\citealt{djoda}; \citealt{dressler}) that relates the velocity
dispersion, effective radius and surface brightness of these sources.
\cite{genzel01} and \cite{tacconi02} concluded that the remnants resemble
moderate mass ellipticals (of stellar mass $\sim$10$^{11}$ \msun). Our new 
data increase the number of objects in the ULIRG luminosity range to be 
compared with early-type galaxies.

\begin{figure*}
\centering
\includegraphics[width=15.4cm]{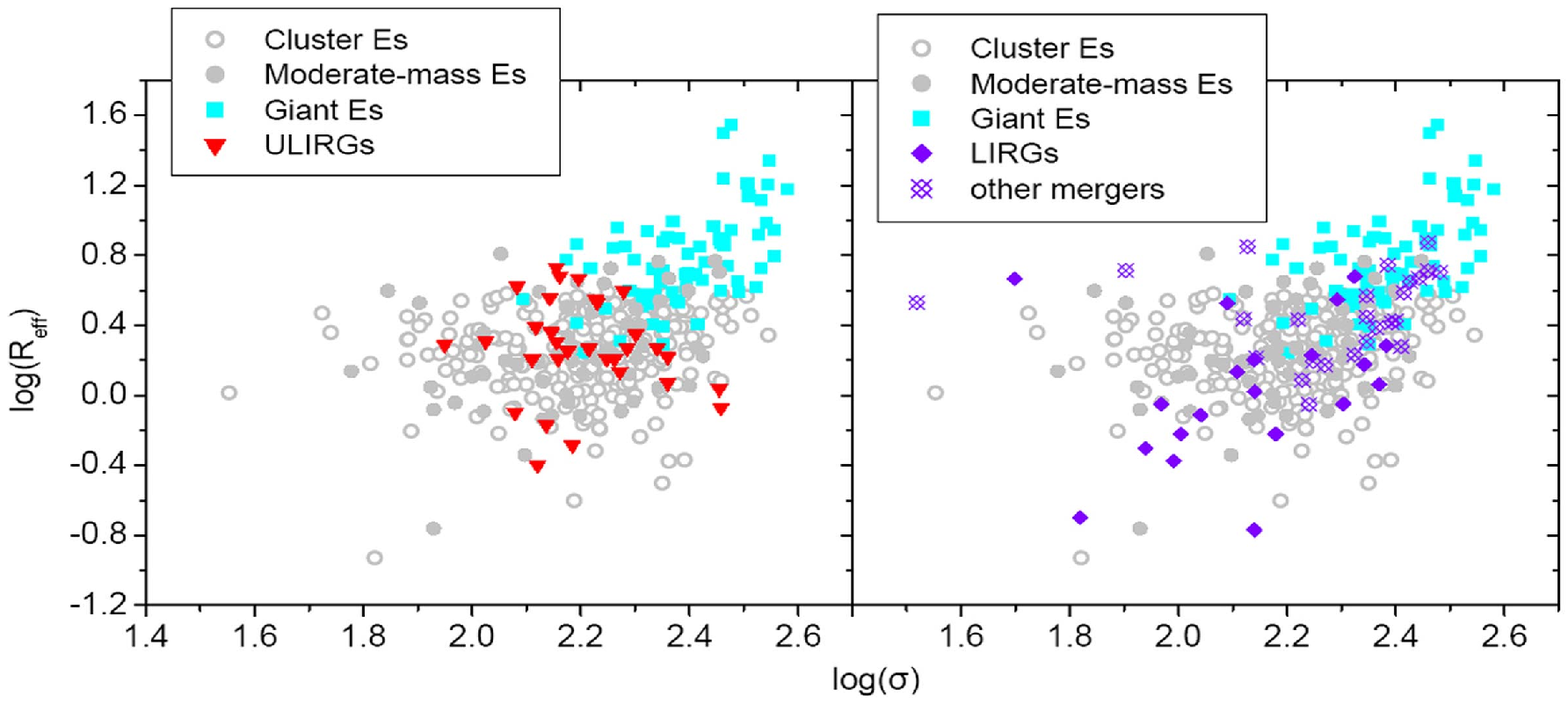}
\caption{
The \reff -$\sigma$ projection of the early-type galaxy fundamental plane. 
The data for the giant boxy and moderate-mass disky Es (squares and 
circles respectively) are taken from \cite{bender92} and \cite{faber97}. More 
(cluster) Es (open circles) are from \cite{pahre}. For viewing clarity,
the various types of mergers are plotted in separate panels. The ULIRG
remnants (29 from this study and 2 from \citealt{rothberg}) are plotted as 
triangles (left panel). LIRGs (diamonds) from \cite{shier}, \cite{james},
\cite{rothberg}, and \cite{hinz} and other (visually-selected) mergers from 
\cite{rothberg} are plotted in diamonds and open-crossed diamonds respectively 
(right panel). The effective radii of all merger remnants used in this 
figure are equal to the averages of their NIR measurements, if more than one 
is available.
\label{fig:fpp}}
\end{figure*}

The $R_{{\rm eff}}$ - $\sigma$ projection of the plane that we construct
from our data is shown in Fig.~\ref{fig:fpp}. Data for early-type galaxies
are taken from \cite{bender92}, \cite{faber97}, and \cite{pahre}. 
Giant boxy ellipticals (squares) occupy the upper-right corner of the FP
projection, while disky, moderate-mass ellipticals (circles) are located at
the center. On the left panel of Fig.~\ref{fig:fpp} we overplot the ULIRG
remnants in triangles; 29 are from this study and 2 from \cite{rothberg}, 
UGC 5101\footnote{The $H$- and $K$-band effective radii of this source 
equal 0.27 kpc and 1.42 kpc respectively (\citealt{sco00}; 
\citealt{rothberg04}).} and AM 2246-490 The location of ULIRGs on 
the plane agrees very well with that of moderate-mass ellipticals.
On the right panel of Fig.~\ref{fig:fpp}, we show where
remnants of other populations are located. Local LIRGs (filled diamonds)
lie from the lower-left corner of the FP projection up to the locus of giant
Es (data from \citealt{shier}, \citealt{james}, \citealt{rothberg}, and 
\citealt{hinz}\footnote{Whenever several velocity dispersion and effective
radius measurements exist in the literature for sources that are not included 
in our sample, we use the average value of those measurements.}.
The position of ULIRGs on this FP projection is more tightly restricted than
that of LIRGs; this result implies that ULIRGs have a narrower range of
intrinsic dynamical properties than LIRGs. Other visually-selected merger
remnants (i.e. sources with perturbed morphology but no IR excess;
\citealt{rothberg}) are shown as open-crossed diamonds. That these merger
remnants lie closer than both the ULIRGs and the LIRGs to the locus of giant
Es possibly reflects a different merger origin (see \S~\ref{sec:disc}).

In the 3-dimensional view of the plane, ULIRGs are known to deviate from
the position of early-type galaxies along the surface brightness axis
due to extinction and population effects  (e.g., \citealt{genzel01}). Due
to the strong starbursts ULIRGs undergo, they have a significant population
of newly formed asymptotic-giant-branch stars, red supergiants and giants.
Thus, their NIR surface brightnesses are higher than those of quiescent
ellipticals (\citealt{pahre}; \citealt{veilleux02}). Prior to comparing
ULIRGs and Es, the removal of light originating from starburst components
is therefore instructive. Since a significant fraction of the starburst
emission is nucleated, we opt to remove the central PSF (\citealt{veilleux06})
simultaneously removing any AGN-originating emission.

In the $H$-band, the nuclear PSF removal has already been performed by
\cite{surace99}, \cite{colina01} and \cite{veilleux06} for most of
the sources in our sample.  For the sources taken from \cite{sco00},
we use the mean $H$-band ratio between the luminosity of the
PSF-subtracted galaxy and the total luminosity, which equals 0.64
(\citealt{surace99}; \citealt{colina01}; \citealt{veilleux06}). The total
magnitudes\footnote{The total magnitude here includes flux integrated up to
the galaxy's truncation radius.} of approximately half of the sources we use 
to construct the 3-dimensional view of the $K$-band fundamental plane are
from \cite{kim02}. For the remaining sources, the data are from \cite{duc97},
\cite{surace99}, \cite{rigopoulou99}, and \cite{sco00}. In the $K$ band,
the value of the ratio between the luminosity of the PSF-subtracted galaxy
and the total luminosity is 0.75 times that in the $H$ band 
(\citealt{surace99}) i.e., 0.48. We compute the mean surface 
brightness (within the effective radius) of each galaxy, $<\mu_{\rm gal}>$, by 
adding $2.5 log(2 \pi R_{\rm eff}^2)$ mags (where $R_{{\rm eff}}$ is in 
\arcsec) to its PSF-subtracted magnitude in both bands. In 
Table~\ref{tab:structure} we tabulate the $H$ and the $K$-band value of 
$<\mu_{\rm gal}>$ for each source.

\begin{figure*}
\centering
\includegraphics[width=12cm]{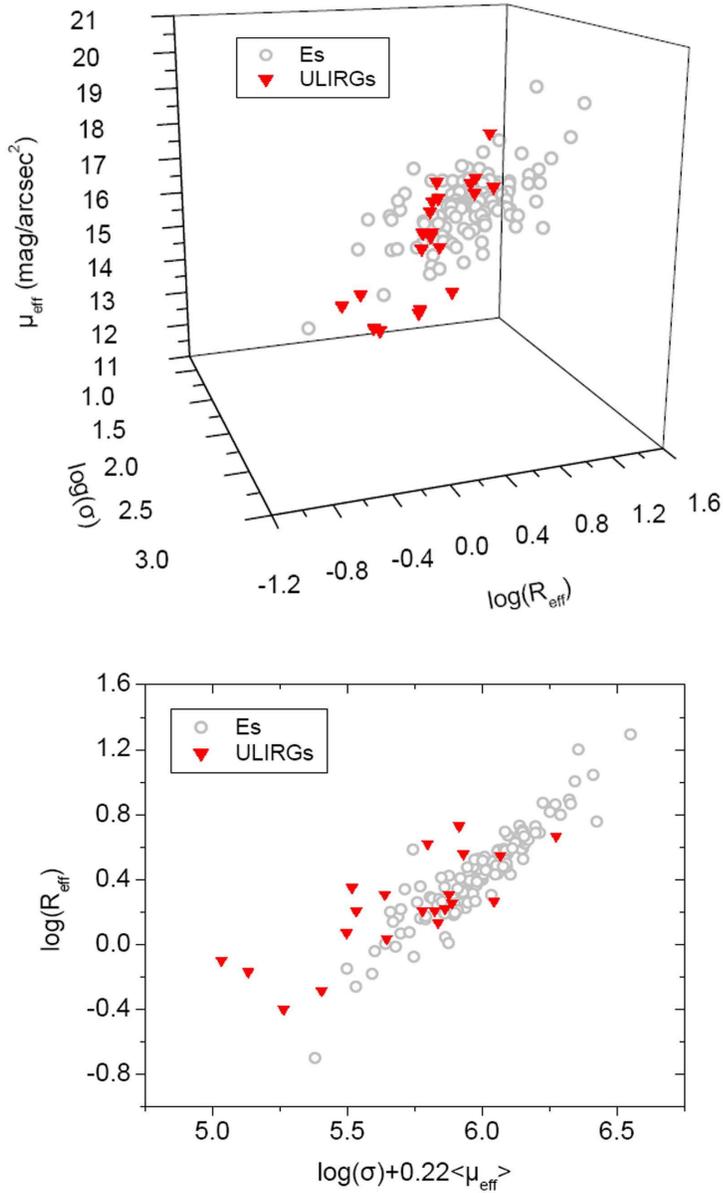}
\caption{
Upper panel: The 3-dimensional view of the $H$-band fundamental plane of 
early-type galaxies. Local early-type galaxies (mainly from the Coma and
Virgo clusters) are plotted as open circles (\citealt{zibe}). 21 ULIRGs 
from this study are plotted as triangles. All data are PSF removed (see 
Table~\ref{tab:structure}). No extinction corrections have been applied 
to the fluxes. The effective radii used are as in Fig.~\ref{fig:fpp}
Lower panel: The fundamental plane, viewed as in \cite{pahre}.
\label{fig:fph}}
\end{figure*}

\begin{figure*}
\centering
\includegraphics[width=12cm]{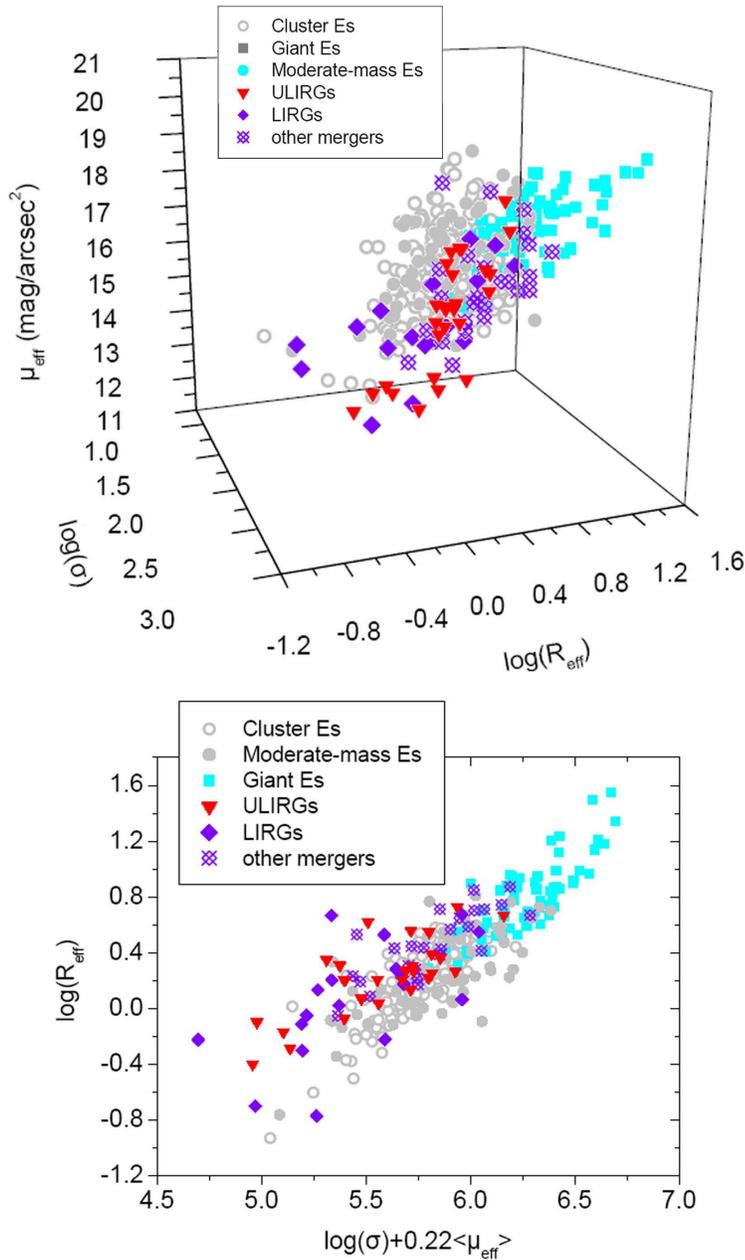}
\caption{
Upper panel: The 3-dimensional view of the $K$-band fundamental plane of 
early-type galaxies. The symbols and effective radii used are identical to 
those in 
Fig.~\ref{fig:fpp}. A PSF estimate has been removed to all photometric data
initially taken from \cite{kim02} (see Table~\ref{tab:structure}). No 
extinction corrections have been applied to the data. From the ULIRGs
plotted in this figure, 25 are from this study and 2 from 
\cite{rothberg}.
Lower panel: The fundamental plane, viewed as in \cite{pahre}.
\label{fig:fp}}
\end{figure*}

In Fig.~\ref{fig:fph} (upper panel) we show the 3-dimensional view of the
fundamental plane in the $H$-band for the 21 remnants of 
Table~\ref{tab:structure}
with a $<\mu_{\rm gal}(H-band)>$ measurement. For viewing clarity, we also 
plot in Fig.~\ref{fig:fph} (lower panel) the fundamental plane in one of 
the representations introduced in \cite{pahre}.
$H$-band data for (mainly cluster) Es are from \cite{zibe} and references
therein. The ellipticals are plotted as open circles\footnote{To enhance the
clarity of the diagram we have excluded the dwarf Es, since they do not
fall on the plane of all other Es.} and the ULIRGs as triangles. In the
$H$-band, the individual PSF subtraction brings ULIRGs close to the
fundamental plane, indicating that ULIRG remnants will resemble Es once
their on-going starbursts cease. That ULIRGs are on average somewhat brighter
than Es is justified by the fact that a (diffuse) part of their starburst
emission remains unremoved after the PSF subtraction. In Fig.~\ref{fig:fp}
we show the $K$-band 3-dimensional view and \cite{pahre} visualization of the
plane for 25 remnants in this study with a $<\mu_{\rm gal}(K-band)>$ 
measurement and two sources from \cite{rothberg}. In the $K$-band, ULIRGs 
lie again close to moderate-mass Es, after the
statistical PSF removal. Along the photometric axis of the plane, 
PSF-subtracted ULIRGs are also close to LIRGs and other merger remnants.

We have not attempted to correct the photometric data of any wavelength
for extinction effects; the value of the usually-assumed equivalent screen 
correction (\citealt{sco00}; \citealt{genzel01}; \citealt{tacconi02}) is 
uncertain and changes after removal of the nuclear PSF. The application of 
an extinction correction will make the surface brightness of ULIRGs brighter
and require more fading for them to evolve onto the fundamental plane.
No k-corrections have been applied since they are negligible for the majority 
of the sources in our sample.

\subsection{Discussion: The role of ULIRGs in the formation of local Es}
\label{sec:disc}

Our main conclusion from the fundamental plane analysis is that ULIRGs 
typically lie on the locus of local moderate-mass ellipticals whose
stellar mass is $10^{10}-10^{11}$ \msun. To further investigate the 
type of elliptical galaxy that ULIRGs form at the highest rates, we 
compare the number densities of local ULIRG remnants and ellipticals as a 
function of their stellar velocity dispersion.

\begin{figure*}
\centering
\includegraphics[width=7cm,angle=270]{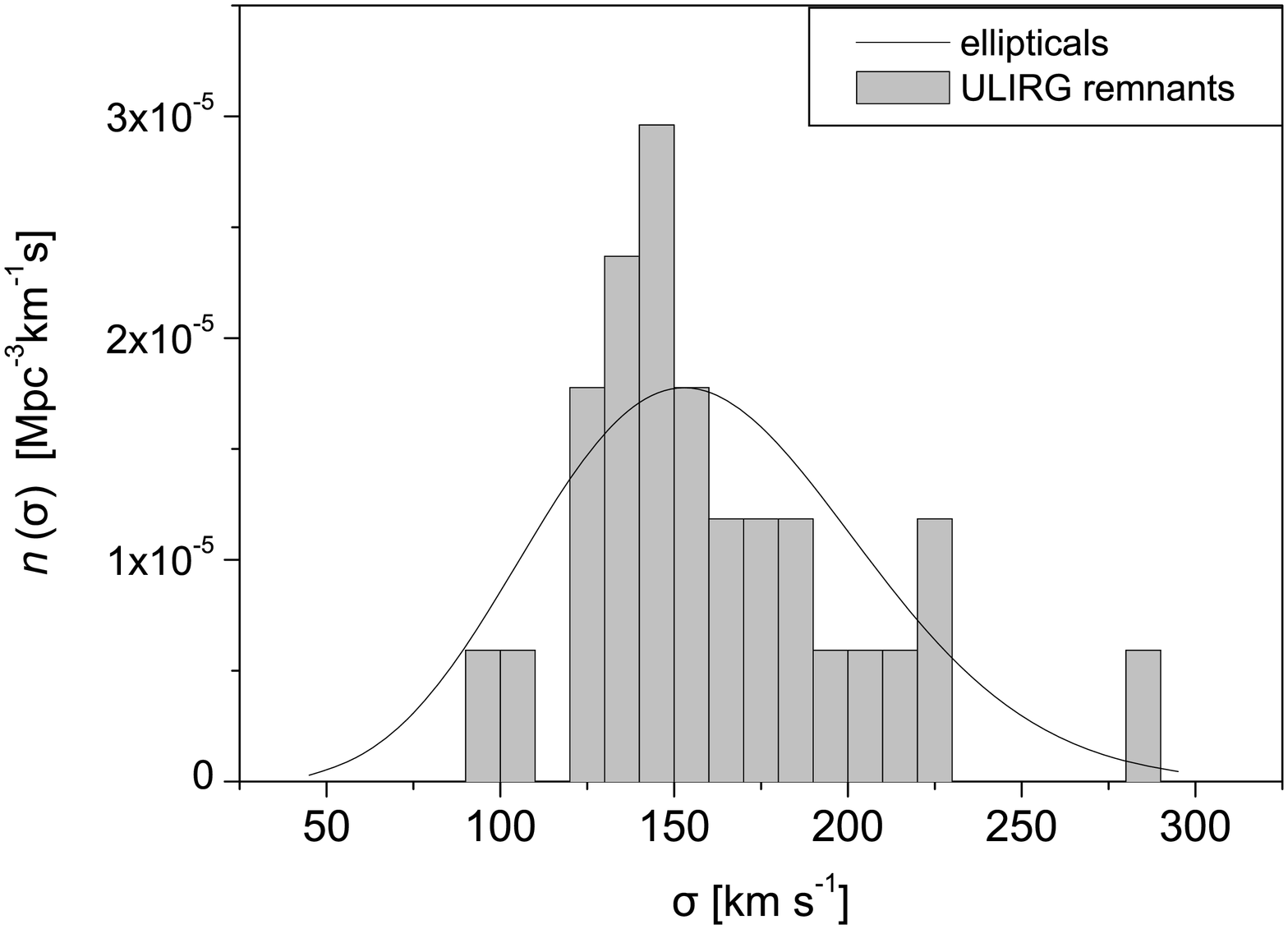}
\caption{
The number density of sources as a function of their stellar velocity 
dispersion is plotted in this figure. The solid line corresponds to the 
number density per $\sigma$ of SDSS ellipticals, computed from their 
dispersion function (\citealt{sheth03}). The number density per $\sigma$ of 
ULIRG remnants, plotted as a histogram, is calculated by multiplying the 
\% fraction of our remnants that resides in each $\sigma$ bin with the local
volume density of ULIRGs from \cite{sanders03}. To facilitate the comparison, 
the ULIRG histogram is further normalized so that its mean has the same 
number density as that of the elliptical $n{(\sigma)}$ distribution.  
\label{fig:lumfun} }
\end{figure*}

In Fig.~\ref{fig:lumfun}, we plot the source number density per stellar 
dispersion in the range between 50 and 300\kms; ULIRGs are plotted as a 
histogram and local SDSS ellipticals ($0.01<z<0.3$; \citealt{bernardi})
as a solid line. For the ellipticals, $n{(\sigma)}$ is computed from the 
velocity dispersion function of \cite{sheth03}. The $n{\rm (\sigma)}$ 
distribution of ULIRGs is calculated by multiplying the \% fraction of ULIRG 
remnants per $\sigma$ bin with the volume density of ULIRGs 
($2.5\times 10^{-7}$ Mpc$^{-3}$, for our cosmology) from \cite{sanders03}. To 
facilitate the comparison between the two $n{(\sigma)}$ distributions, we 
apply a normalization factor of $7\times 10^{3}$ to the ULIRG histogram so 
that its mean has the same number density as that of the SDSS ellipticals.
Physically, this normalization factor is related to the ratio between
the time over which ellipticals have been formed and the lifetime of a
single ultraluminous burst. Its value is only a rough estimate  
since it depends on the completeness of the 1 Jy and the SDSS (sub-)samples 
and the fact that local ultraluminous mergers are not the only mechanism to 
produce elliptical galaxies.

The $n{(\sigma)}$ distribution of SDSS elliptical galaxies has a mean 
dispersion similar to that of ULIRGs, 157 (with a standard deviation 
of 54) \kms. In other words, the descendants of ULIRGs will resemble the 
ellipticals that are most common in the local Universe. Local ULIRGs seem to 
form at highest rates sources of characteristic $\sigma$ between 130 and 160 
\kms. That local Es are mainly produced by mergers of spirals of specific 
luminosities (or masses) is also found in the recent simulations of
\cite{kavi}.

At low dispersions, the ratio between $n{(\sigma)}$ and $n{(<\sigma>)}$ seems 
to be lower for ULIRGs than for ellipticals. This deviation can probably 
be attributed to the fact that mergers of gas-rich galaxies below a certain 
mass threshold do not possess enough gas to undergo a ULIRG phase.

Deviations between the two distributions also exist at the high-dispersion 
end. For most bins at the high-dispersion end, the ULIRG distribution seems
again underpopulated compared to that of SDSS ellipticals. We cannot formally 
rule out the possibility that the most massive ULIRGs have been missed 
due to sample incompleteness, small volume densities, or possibly
shorter (than the average) burst timescales for the most massive sources. 
However, one source with high velocity dispersion, Mrk 273, has been 
observed in our sample (see also \citealt{hinz}). Even for Mrk 273, the total 
mass of the system 
(which also depends on its effective radius) is insufficient to classify it 
as a giant elliptical (of stellar mass $>5\times 10^{11}$\msun). Moreover, the 
fundamental plane analysis for all ULIRGs in our sample seems to suggest that 
ULIRGs do not account for the formation of a significant fraction of the 
giant elliptical galaxies in the local Universe. To quantify this statement, 
we apply a KS test to the ULIRG remnants and the (literature-selected) giant 
Es we used to construct the fundamental plane, acknowledging the different 
space-volumes they occupy. We find that the probability that ULIRG remnants 
and the giant ellipticals originate from the same population is extremely 
small,  
10$^{-9}$ (see also \citealt{genzel01}); this probability indicates that 
despite the small number statistics, the confidence of this result is very 
high. For their morphologically selected remnant sample, \cite{rothberg} find
a higher, but still insignificant, probability of 10\%.

To further assess how typicially ULIRGs form giant Es, we used the mass 
function of local Es to estimate what fraction $f_{\rm giant}$ of the 
sources with mass $>10^{10}$\msun\ (mass threshold similar to that of ULIRGs) 
correspond to giant ellipticals (of stellar mass $>5\times 10^{11}$\msun).
Given that the fraction may significantly vary according to the adopted mass 
function, we used the best fit to the $K$-band mass function from various 
authors in the literature. Giant ellipticals comprise 18.7\% of local Es with 
mass $>10^{10}$\msun\ for the mass function that \cite{bell03} derived for Es
in the Sloan Digital Sky (SDSS) and the Two Micron All Sky (2MASS) surveys 
assuming a diet\footnote{In this initial mass function, the number of 
low-mass stars is reduced leading to a total stellar mass reduced by 30\%.} 
Salpeter initial mass function (IMF). The fraction $f_{\rm giant}$ is 20.1\% 
for the 2MASS $K$-band luminosity function of local Es (\citealt{kochanek01}), 
which we convert to a mass function using $m/L=1.32$ \mlr\ (\citealt{cole01}).
This $K$-band $m/L$ ratio is computed for both early- and late- type 
galaxies in the 2MASS and 2dF Galaxy Redshift Surveys, assuming a Salpeter 
IMF. An estimate of the $m/L$ ratio for local ellipticals only can be derived 
from the work of \cite{cappellari05} using the mean $I$-band $m/L$ value 
and I-K color of the SAURON project sources. Combining the 
2MASS $K$-band luminosity function with this early-type-galaxy $m/L$ value
(0.75 \mlr) yields $f_{\rm giant}$=10.9\%. Although $f_{\rm giant}$ is between 
10\% and 20\%, we do not observe any source of $m>5 \times 10^{11}$ \msun\ 
in our sample of 29 ULIRG remnants. This fact indicates that 
local giant ellipticals cannot be accounted for, if assumed to originate 
only from local ultraluminous mergers.

One possible scenario for the formation of local giant Es is that most 
of these objects have formed at higher redshifts (e.g. \citealt{swinbank}). 
This scenario is based on the fact that the more massive early-type
galaxies are, the faster and the earlier their
stellar populations have formed (\citealt{thomas05}; \citealt{vandokkum} and
references therein). It is reinforced by the argument that local giant Es are 
mainly located in clusters whereas ULIRGs lie in the field (\citealt{tacconi02}).
Another likely scenario, which may hold simultaneously, is that other types 
of mergers (e.g. E-S or E-E ones) account for the formation of $\sim$60\% 
of elliptical galaxies (\citealt{khoch03}; \citealt{kavi}) at low redshift.
E-S mergers enhance the star-formation activity already existing in one of 
the merging components and, for adequate amounts of gas, may appear as LIRGs. 
Gas-depleted E-E mergers (also known as ``dry'' mergers) are not easy to 
detect observationally (e.g, \citealt{vandokkum}; \citealt{bell06}) since 
they do not have other 
unambiguous signatures such as luminous/ultraluminous phase(s) beyond their
perturbed morphologies; therefore, they can mainly be traced by visual 
identification. Dry merging is now believed to be a key element in the 
formation of local giant Es (\citealt{naab06a}) and is a likely explanation 
for the locations of some optically selected mergers in the fundamental 
plane (see \S~\ref{subsec:fp}).

\cite{rothberg} recently compiled a sample of local merger remnants selected 
by visual inspection. Such a sample probably comprises merger remnants of all 
possible (S-S; E-S; E-E) categories. \cite{rothberg} measured $\sigma$ from 
Ca II triplet (optical) spectroscopy for 38 remnants, of which, 10 were 
LIRGs and 2 were ULIRGs. They found that the hypothesis that the
(U)LIRGs in their sample and giant ellipticals can be 
drawn from the same parent population has a probability of 10\%. 
\cite{lake} also obtained optical velocity dispersion measurements for
13 merger remnants. Of those, 10 have Infrared Astronomical Satellite (IRAS)
fluxes; none is a ULIRG and only one source, 
AM 0921-225, is a LIRG\footnote{We computed the IR luminosity of these sources
with the formula of \cite{sami96} using mid-infrared fluxes from \cite{moshir},
\cite{iras88}, and the NASA/IPAC Extragalactic Database}. 
\cite{rothberg} computed the probability that the remnants of \cite{lake} 
and giant Es originate from the same parent population and found it  
to be 5\%. 

The conclusion from the comparison of the optically-selected 
remnants of \cite{rothberg} and \cite{lake} with the IR-selected remnants in 
our study (and those of \citealt{genzel01} and \citealt{tacconi02}) is what 
is expected from a theoretical point of view: while mergers account for the 
formation of galaxies of various morphological classes and mass contents 
(e.g., \citealt{schweizer}; \citealt{springelspiral}; \citealt{springelred}), 
ultraluminous IR mergers seem to have a specific output (moderate-mass 
ellipticals) originating from their specific input (mostly sub-m$_*$ spirals).

\subsection{Wavelength Dependence Of The Stellar Velocity Dispersion
Measurement}

Systematic differences between the measured and the actual values of the 
host dispersion may arise when extracting $\sigma$ from different wavelengths, 
i.e. from the Ca II triplet in the optical (e.g. \citealt{rothberg}) and 
the CO bandheads in the NIR. The host kinematics extracted from 
the NIR bandheads are often representative of young stellar 
populations. These populations could still be linked to the gas from
which they formed (which is believed to settle into a disk earlier 
than the stars in the progenitor disks; \citealt{mihos96}), 
and have less perturbed orbits than the old stars. 
Host dispersions extracted from the CO bandheads could be 
systematically lower than those of the merging bulges. However, the low 
\vrot$/\sigma$ ratio that we measure for our ULIRGs is a good 
indication that even the young stellar populations are significantly 
heated, and therefore, systematics originating from NIR dispersion 
measurement are unlikely to have a major effect on our conclusions.

In the case of the Ca triplet, systematic errors are mainly related to 
the presence of dust. Due to scattering of light from dust particles, photons 
originating from high-velocity stars in the center of the galaxy will be 
scattered into random lines of sight, biasing the LOS velocity distribution 
(\citealt{baes02}). The effects of dust in the observed stellar kinematics 
depend upon the dust mass and extent (\citealt{baes02}). Since ULIRGs are 
highly obscured systems, the stellar dispersions extracted from the Ca triplet 
may be systematically higher than their intrinsic values. 

Such possible systematics have not been investigated in ULIRGs.
\cite{silge} have attempted to quantify the discrepancy between 
dispersion estimates derived from the NIR and the optical 
regimes for local quiescent galaxies. While they found no 
significant difference for elliptical galaxies, the systematics were 
non-negligible in S0 lenticulars, with the largest difference in $\sigma$ 
($\sim$30-40\%) found in the most massive sources of their sample. 
The opposite conclusion is found from the velocity dispersion
of ellipticals and Seyfert-type AGNs in \cite{oliva95} and \cite{oliva99}.
The $\sigma$ values for the objects in their sample were systematically 
greater in the NIR than in the optical. The difference was large for type 1 
Seyferts (up to a factor of 2) but low for type 2 Seyferts and ellipticals
($<$10\%). However, one of the uncertainties often underestimated in 
these comparisons is that the spectral resolution in the NIR was 
significantly lower than that in the optical. For \cite{silge}, $R$ was
2300 in the $K-$band; for Oliva et al. (1995; 1999) it was equal to 1900 
and 2500 in the $H-$ and in the $K-$band respectively. To get a better 
estimate of the possible systematic errors in $\sigma$, we compare the 
results for LIRGs with a Ca II triplet $\sigma$ measurement 
(\citealt{rothberg}) that also have CO observations of relatively high 
spectral resolution (e.g., equal or better than 
ours). These sources (NGC 1614, NGC 2623, NGC 4194, Arp 193, and
IRAS 20551-4250) have been observed in the NIR by \cite{james}, 
\cite{genzel01}, \cite{rothberg}, and more recently, \cite{hinz}. For 
most sources, the optical dispersions were higher than the NIR ones, 
but the systematic difference was small (7\% on average). 
On the other hand, significant deviations (of a factor of 2) exist between 
the NIR results of \cite{james} and \cite{hinz} and the older (also NIR
but lower-resolution) data of \cite{shier} for NGC 1614 and NGC 2623. 
It is therefore not clear whether the measurement of $\sigma$ from 
the different features would systematically yield different results 
for ULIRGs, and if yes, whether they would reflect real dynamic trends 
instead of measurement errors. Errors may also be easily introduced from 
the use of different stellar templates by various authors.

We investigate whether the use of the Ca II triplet for the 
measurement of $\sigma$ may affect our results on the formation of (giant) 
Es under the hypothesis that the optical $\sigma$ values can be indeed 
greater than the NIR ones due to population and extinction effects. We compute 
the maximum possible increase that can be applied to the dispersions of the 
sources in our sample, which is the maximum reported deviation (40\%) in the 
results of \cite{silge}, corresponding to an addition of $\sim$60 
\kms\ in $\sigma$ or a shift of 0.15 in the (logarithmic) horizontal axis of 
the fundamental plane. It is clear that this shift brings merger remnants 
closer to, but not into, the region populated by giant ellipticals. We also 
recompute the masses of the individual remnants in our sample after increasing 
their dispersion by 40\%. The average mass is now 1.69$\times 10^{11}$ \msun, 
which is again close to $m_*$ and there is still no source more massive than
5$\times 10^{11}$ \msun. Therefore, our main conclusions are 
insensitive to the wavelength dependence of the stellar velocity dispersion 
measurement. Furthermore, the \vrot (obs)$/\sigma$ ratios\footnote{not 
corrected for inclination effects} of S0 lenticulars are typically greater than
those of ULIRGs; e.g, for the combined sample of \cite{pizz} and \cite{cap05b} 
the ratio is twice that of the remnants in our sample (0.6) indicating that 
deviations of the order 30-40\% between the CO and Ca dispersion measurement 
are probably high for ULIRGs.


\section{Black Holes in ULIRGs}
\label{sec:black}

\subsection{A Picture Of The \mbh-$\sigma$ Relation Time Evolution}
\label{sec:theory}

The size of a black hole seems to be closely linked to the depth of the 
potential well in which it forms and grows (\citealt{feme00}). 
This is reflected, e.g., in the local \msigma\ relation, the correlation 
between the black hole mass and the stellar dispersion in the
bulge of the host galaxy (\citealt{feme00}; \citealt{gebhardt}). Since the 
\msigma\ relation is found for virialized bulges, it should be valid 
at the end of the merger process, when the bulge stellar kinematics have 
reached their final dynamical state and AGN winds and supernovae feedback 
have expelled the gas away from the nucleus, preventing further BH growth, 
terminating the starburst phase, and making the system resemble an elliptical 
galaxy. The merger remnants in recent simulations (\citealt{diMatteo05};
\citealt{springel05}; \citealt{robertson05}) are able to reproduce the 
\msigma\ relation by subjecting a portion of their interstellar gas to 
accretion and feedback.

For ULIRGs however, the conversion of the host velocity dispersion into \mbh\ 
carries the uncertainty of applying the \msigma\ relation to systems out of 
dynamical equilibrium. It is not yet known if, or to what extent, the \msigma\
relation is valid between the first encounter and shortly ($\lesssim 10^8$ 
yrs) after the final coalescence, when most of the ultraluminous infrared 
activity occurs. 

To investigate whether (and under which conditions) merging disk galaxies 
fall on the \msigma\ relation during interaction phases prior to coalescence, 
we ran gas-rich merger simulations (details presented in \citealt{naab06}) 
that have already been discussed briefly in \pone. To include the 
effects of a dissipative component we replaced 10\% of the stellar mass in the 
initial disks with isothermal gas at a temperature of approximately 10$^4$K.  
The initial scale length of the gas disk was equal to that of the stellar 
disk, $h$. Each galaxy had a stellar bulge with 1/3 of the disk mass and was 
embedded in a pseudo-isothermal halo to guarantee a flat rotation curve at 
large radii. All galaxies approached each other on nearly parabolic orbits 
with a pericenter distance of two disk scale lengths. The spin orientations
were both prograde and retrograde. The evolution of the 
stars and the gas was computed with the N-body smoothed-particle-hydrodynamics 
code VINE using an isothermal equation of state for the gas. 
For this study we analyzed mergers with 16 different initial disk orientations 
(geometries 1-16 of \citealt{naab03}) and mass ratios 1:1 and 3:1.

We analyzed snapshots in the orbital plane approximately every half-mass 
rotation period of the more massive disk to follow the time evolution of the 
projected central stellar velocity dispersion and the gas accretion history 
onto the center of the system. At each snapshot, we calculated \mbh\ from 
the velocity dispersion assuming that the \msigma\ relation is always valid.
As in \pone, we used the \cite{tremaine02} formula
\begin{equation}
\mbox{$M_{BH}=1.35 \times 10^{8} [\sigma/ 200]^{4.02}$\msun,}
\end{equation} 
where $\sigma$ is in units of \kms, to calculate the black hole masses.
To quantify the gas accretion, we computed the total gas mass $M_{\rm gas}$
that has ever reached a radius of $0.1h$ (i.e. twice the resolution element of 
the simulations)  relative to the mass center of the system. By 
calculating the ratio between the quantities \mbh\ and 
$M_{\rm gas}$ we derive the accretion efficiency $\epsilon$ that is needed
to maintain the black hole mass on the \cite{tremaine02} formula at any time. 
We define accretion efficiency $\epsilon$ as the fraction of the gas that 
accretes onto the black hole from the gas accumulated into the nuclear 
region.

\begin{figure*}
\centering
\includegraphics[height=11cm,width=14cm]{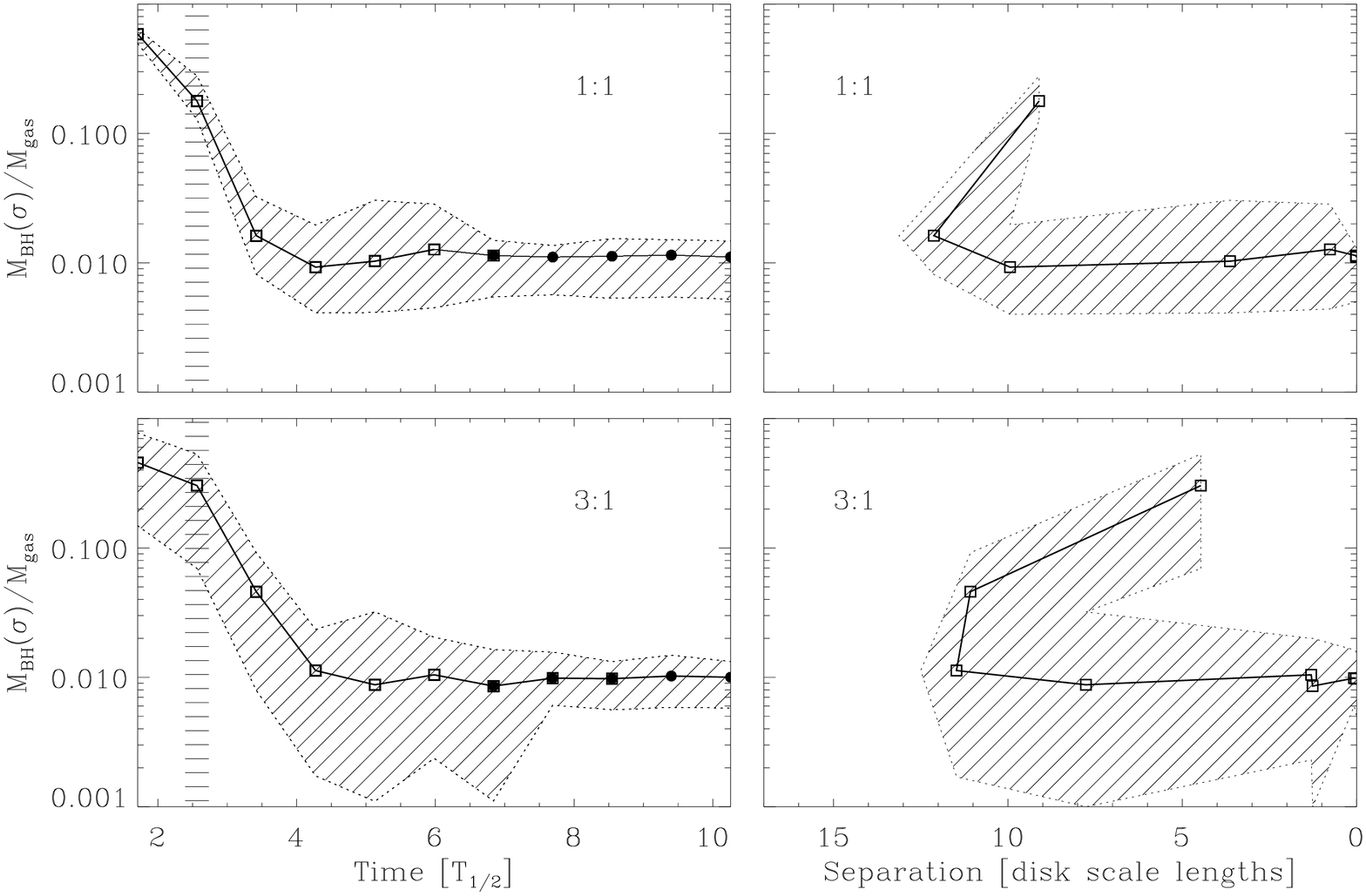}
\caption{
Evolution, during a disk-galaxy merger, of the average black-hole mass 
inferred from the central (line-of-sight) stellar velocity dispersion of the 
simulated galaxies. The black hole mass is in units of the total 
gas mass accreted onto the center of the simulations.
In the left column, the evolution is given as a function of time. 
Time is plotted in units of the half-mass rotation period, $T_{1/2}$, of 
the more massive progenitor disk. Open squares represent disks that are still 
separated, filled dots indicate fully merged systems. The spread originating 
from the initial disk orientations is indicated by the diagonally shaded
area. The vertically shaded area indicates the first data point after 
the first encounter. In the right column, the evolution is plotted as a 
function of nuclear separation of the interacting galaxies. The nuclear 
separation unit is the disk scale length $h$ of the more massive disk.  
We begin plotting data points for 1 step prior to first encounter, as in 
the left panels. We show 1:1 mergers in the upper and 3:1 mergers in the 
lower panels.
\label{fig:modelt}}
\end{figure*}

In Fig.~\ref{fig:modelt} we show the evolution of the accretion efficiency
(needed to keep the black hole on the \msigma\ relation) as a function of 
time (left column) and nuclear separation (right column). Time is given 
in units of the half-mass rotation period, $T_{1/2}$, of the more massive 
progenitor disk. The nuclear separation is given in disk scale lengths of the 
more massive disk. The models are scale-free; scaling the parameters to e.g., 
those of the Milky Way corresponds to a unit time of 13 Myrs and 
a unit length of 3.5 kpc. Scaling them to the mean scale length and 
rotational velocity of all individual progenitors in our binary ULIRG 
sample yields a unit time of 13 Myrs and a unit length of 1.3 kpc. The 
efficiency is averaged over the 16 initial geometries for 1:1 and 3:1 
mergers (upper and lower panels). The diagonally shaded area indicates the 
spread due to the varying initial disk geometries. 

According to the left panels of Fig.~\ref{fig:modelt}, before nuclear 
coalescence, the estimated black hole mass and the gas accumulated at the 
center of the simulation already scale linearly. In other words, if we assume
that the \msigma\ relation is valid at {\it any} time during the merger, then 
we find that the accretion efficiency remains constant shortly after the first 
encounter (shown in vertical dashes) until relaxation. Vice-versa, if we 
assume that $\epsilon$ stays constant during the merger and require only
that the \msigma\ relation be valid {\it at relaxation}\footnote{This 
assumption is reasonable since ULIRGs will eventually form Es.}, we find 
that the black hole mass can be computed from the \msigma\ relation from 
midway of the first encounter through relaxation. 

The implication of these simulations is that the \msigma\ relation can be 
used to compute black hole masses in merging systems, from mid-way between 
first encounter and nuclear coalescence until the remnant becomes an 
elliptical galaxy, as long as $\epsilon$ remains constant. This is true
for both equal- (1:1) and unequal- (3:1) mass major mergers. A constant 
efficiency $\epsilon$ conceptually corresponds to subjecting a specific 
gas fraction to AGN feeding, an assumption also made for models that include 
star formation and ISM feedback (e.g. \citealt{springel05}). However, it is
likely that on short timescales (e.g. during specific accretion events) 
the efficiency $\epsilon$ may vary significantly. Therefore individual black 
hole estimates calculated from the \msigma\ relation carry this uncertainty.

In the right panels of Fig.~\ref{fig:modelt},
the ``folding'' of the efficiency $\epsilon$ at large nuclear separations is 
due to the fact that the galaxies move towards apogalacticon before they 
fall back together. Clearly, at these merger phases, the use of the \msigma\ 
relation is misleading. For a constant $\epsilon$, the \msigma\ relation 
begins to be valid beyond 5 disk scale lengths, which corresponds to a nuclear 
separation $\sim$7 kpc for our ULIRGs (see \pone). If the assumption of a
constant $\epsilon$ holds, the black hole estimates for $\sim2/3$ of the 
binary sources may be considered reasonable (but still lower limits on 
their final values). For the ULIRG 
remnants, computing BH masses from stellar dispersions seems also plausible. 
This is further supported by the fact that the kinematics of the latter are 
expected to have almost settled to their relaxation values 
(\citealt{mihos00}).


\subsection{Black Hole Sizes and Accretion Rates}
\label{sec:bhm}

We calculate the individual black hole masses of the sources in our sample 
using the \cite{tremaine02} formula and tabulate the results in 
Table~\ref{tab:velocities}. The mean black hole mass of the ULIRG 
remnants is $8.4 \times 10^7$\msun.
Since our sources are still in an ultraluminous IR emission phase, gas and 
dust are still present in the nuclear region and will continue accreting 
onto the black hole. However, beyond coalescence, the timescales for further 
gas accretion onto the black hole are probably shorter than those of the 
pre-coalescence starbursts (e.g. \citealt{springel05}). Therefore, the black 
hole masses of the fully relaxed systems will be somewhat larger than 
those presented in Table~\ref{tab:velocities}. Simple gas content (typical 
gas mass of ULIRGs is $5 \times 10^9$ \msun; \citealt{dowso98}) and accretion 
efficiency (0.01, see \S~\ref{sec:theory}) arguments indicate that the 
additional increase of the remnant BH masses will not exceed 
10$^8$ \msun\ by the time the ultraluminous activity ends. In this 
order-of-magnitude calculation we have ignored the fact that only part of the 
gas reaches (or remains in) the center of the merging system, further reducing 
the upper limit on the final black hole mass.

We calculate the BH 
mass each source would have, if it were accreting at the Eddington rate
\begin{equation}
\mbox{$L_{\rm Edd} / L_{\odot} = 3.8 \times 10^{4} M_{\rm Edd} / M_{\odot}$.}
\end{equation} 
To calculate the Eddington BH mass $M_{\rm Edd}$, we assign to the Eddington 
luminosity $L_{\rm Edd}$ half of the luminosity emitted in the IR, given that 
some ULIRGs are largely starburst- while others are AGN- powered 
(\citealt{genzel98};  \citealt{rigopoulou99}; \citealt{veilleux99}; 
\citealt{joseph99}; \citealt{sanders99}). The Eddington efficiency 
$\eta_{\rm Edd}$, the ratio of the Eddington \mbh\ estimate over the dynamical 
black hole mass, is given in Table~\ref{tab:velocities}. The mean Eddington 
efficiency of the merged-ULIRG sample is 0.89, with a wide range of 
values. While statistically appropriate, the assumption that 50\% of the IR 
luminosity originates from the AGN may make some sources appear as if accreting
at super-Eddington rates (see Table~\ref{tab:velocities}). If we exclude 
these sources from the statistics and recalculate the average efficiency,
we find it to be 0.39.


The inferred accretion rates may be higher than in reality if an overestimated 
fraction of the IR emission is assigned to the AGN. This could occur if, for 
example, the starburst is still the dominant source of the IR luminosity 
after the merging nuclei coalesce. The relative strengths of the starburst 
(L$_{\rm SB}$) and the AGN (L$_{\rm AGN}$)
luminosity at each merger phase are uncertain. However, the MIR ISO 
spectroscopic study of local ULIRGs by \cite{genzel98} and \cite{rigopoulou99}
indicates that most ULIRGs are starburst dominated systems, implying that
probably less than 50\% of the IR luminosity should be assigned to the
AGN. 
\cite{veilleux02} have found that the strength of nuclear 
continuum emission increases with decreasing nuclear separation. Therefore, 
the Eddington efficiencies of the progenitors presented in \pone\ may be 
systematically overestimated compared to those of the remnants. Future papers
presenting SPITZER MIR spectroscopy of local ULIRGs will indicate the 
appropriate luminosity fraction that needs to be assigned to the AGN for 
ULIRGs before and after nuclear coalescence.
 



\section{Conclusions}
\label{sec:conc}
We have acquired NIR spectroscopic, long-slit, data of 54 ULIRGs 
at a variety of merger phases to trace the evolution of their host 
dynamical properties. From the analysis of the stellar kinematics in
29 ULIRG remnants, we find that:
\begin{enumerate}
\item
Indications of an increase of the stellar random motions exist as the merger 
advances. The mean stellar $\sigma$, as measured from the CO rovibrational 
bandheads, equals 142 \kms\ for the binary sources (of mean nuclear 
separation 9.4 kpc, including IRAS~00456-2904 and IRAS 09111-1007) 
and 161 \kms\ for the remnants (of nuclear separation $<$1.5 kpc). 
This difference in the means of the pre- and post- coalescence 
distributions is marginally significant and requires more data to be accurately
constrained. This increase of the stellar dispersion observed in 
ULIRGs corresponds to only a part of the dynamical heating that occurs during 
the merger, since the merger timescales are longer
than those of the ultraluminous starburst.
\item
The dynamical and structural properties of the remnants indicate that
they originate from mergers mainly of 1:1 and 2:1 progenitor mass ratios.
This confirms what we found in \pone\ by directly measuring the masses of the
individual progenitors of binary ULIRGs. 
\item
Ultraluminous mergers will mainly lead to the formation of moderate mass 
ellipticals (of stellar mass $\sim$10$^{11}$\msun). Depending on the
definition of $m_*$, ULIRGs are between sub- and $\sim m_*$.
ULIRGs are located in a region of the fundamental plane of early-type 
galaxies different from that of local giant Es, indicating
a different formation history for most of the latter. Local mass and 
dispersion functions support this argument; however, the current statistics 
cannot exclude the formation of few giant Es from ULIRGs.
\item
We have performed simulations to investigate whether a black hole mass-host 
dispersion relation can be used to calculate the black hole masses 
of our ULIRGs. We find that already before nuclear coalescence, the mass of 
the gas that falls into the center of the merging system scales 
linearly with the black hole mass predicted by the \msigma\ relation. 
However, this is only true (and conversely) if the efficiency of gas accretion 
onto the BH from its surroundings remains constant with time. 
\item
The black hole masses of the merged ULIRGs are of order 10$^7$-10$^8$\msun\
and their accretion rates are high (Eddington efficiencies often $>0.5$). If 
the AGN luminosity output of a ULIRG nucleus increases with time, our 
accretion rates in pre-coalescence ULIRGs may be overestimated relative to
those in post-coalescence ULIRGs. 
\end{enumerate}








\acknowledgments

We are grateful to N. Scoville for providing $H$-band NICMOS imaging data 
for seven sources and D. C. Kim for giving us 
prior-to-publication NICMOS photometric results for several sources. We 
would like to thank the ESO VLT staff for their support both in the service 
and visitor mode data acquisition. We acknowledge the detailed input of
the anonymous referee that improved the clarity of the manuscript. S. Veilleux 
was supported in part by NASA grant GO-0987501. A. J. Baker acknowledges 
support from the National Radio Astronomy Observatory, which is operated by
Associated Universities, Inc., under cooperative agreement with the National
Science Foundation. 

\newpage


\appendix
\section{APPENDIX: Notes on individual sources}
\label{sec:individual}

IRAS 00091-0738:
This source has two overlapping nuclei separated by 1.4 kpc. Archival Wide 
Field Planetary Camera 2 data of this source (PI Borne) also indicate an 
optical nuclear separation of 1.5 kpc.

IRAS 00456-2904:
According to the imaging analysis of \cite{kim02} and \cite{veilleux06} this 
source is probably a binary system at projected nuclear separation of 22.2
kpc (converted to our cosmology). The redshift of the NE nucleus is not 
spectroscopically confirmed but its ambiguous morphology indicates an 
interaction. The data presented in this paper are for the more luminous 
(late-type-host) SW nucleus.

IRAS 09111-1007: \cite{duc97} find that this source has two widely 
separated interacting nuclei; the ultraluminous IR component is the western 
source. The nuclear separation measured from our acquisition images equals 
40.4 kpc. 

IRAS~14348-1447: 
According to the NICMOS imaging of \cite{sco00}, this source has a projected 
nuclear separation of 6.0 kpc (converted to our cosmology).

IRAS~14378-3651:
Two components separated by 1.5 kpc are visible in the acquisition image
of this source. 

IRAS~15250+3609:
The two components of this source are separated by 0.8 kpc (\citealt{sco00}).

IRAS~17208-0014: \cite{hinz} measure the velocity dispersion of this source
to be 125 $\pm$ 28 \kms.

IRAS~20551-4250: \cite{rothberg} find a dispersion of 185 $\pm 6$ \kms\ 
for this source.

IRAS~23578-5307:
Our acquisition image indicates the presence of two nuclei separated
by 1.4 kpc.

Arp 220:
This source has a projected nuclear separation of 0.3 kpc 
(\citealt{scoville98}).

NGC~6240:
The two nuclei of this source are separated by 1.4 kpc and still show some 
discrete velocity patterns. The velocity dispersion of the system peaks 
between them; at the position of this peak the stellar kinematics may 
reflect localized motions of self-gravitating gas (\citealt{tecza}; 
\citealt{genzel01}). For the velocity dispersion of this source we use 
the average of the values at two nuclei and at the internuclear peak (229 
\kms), which is close to the luminosity-averaged value, 225 \kms\ 
(\citealt{tecza}).

Mrk~273:
This system has two nuclei separated by 0.8 kpc (\citealt{sco00}). 
\cite{hinz} find a stellar velocity dispersion of 232 $\pm$43 \kms\ for
the north component of this source.





\clearpage


\begin{deluxetable}{cccccccc}
\tablecolumns{8}
\tabletypesize{\tiny}
\tablewidth{0pt}
\tablecaption{\label{tab:list} Source List}
\tablehead{
\colhead{Galaxy} & \colhead{RA}   & \colhead{Dec}    & \colhead{$z$} &
\colhead{log($L_{\rm IR}/$\hbox{L$_{\odot}$})} & \colhead{slit P.A.}   & 
\colhead{$t_{\rm integration}$}  & \colhead{Merger phase}\tablenotemark{a}  \\

\colhead{(IRAS)} & \colhead{(2000)}   & \colhead{(2000)}    & \colhead{} &
\colhead{} & \colhead{(\degree)} & \colhead{(mins)} & \colhead{classification} 
}
\startdata
00091$-$0738 & 00:11:43.3 & $-$07:22:08 &  0.118 & 12.19 & 17,106 & 60,60 & R\\
00199-7426 \tablenotemark{b}& 00:22:07.0 & -74:09:42 &  0.096 & 12.23 
& -15,75,74 & 60,60,60 & U \\
00262$+$4251 \tablenotemark{c}
& 00:28:54.0 & $+$43:08:18  &  0.0927 & 12.02 & 45,0  & 20,60 & R \\
00397$-$1312 & 00:42:15.5 & $-$12:56:04 &  0.262 & 12.90 & -1,89 & 120,120 &R\\
00456$-$2904 \tablenotemark{c} 
& 00:48:06.8 & $-$28:48:19 & 0.110 & 12.12 & 30 & 40 & B\\
F01004$-$2237 & 01:02:49.9 & $-$22:21:57 &  0.118 & 12.24 & -1,89  & 60,60 &R\\
01166-0844 \tablenotemark{b} & 01:19:07.6 & -08:29:10 &  0.118 & 12.03 
& -60,29,29 & 60,60,60 &B\\
01388$-$4618 \tablenotemark{c}
& 01:40:55.9 & $-$46:02:53 & 0.090 & 12.03 & 0,90 & 40,40 &R\\
01572$+$0009 (Mrk 1014) \tablenotemark{c}
& 01:59:50.2 & $+$00:23:41 & 0.163 & 12.53 & 20, -70 & 80, 60 &R\\
F02021$-$2103 & 02:04:27.3 & $-$20:49:41 &  0.116 & 12.01 & 53,142 & 60,60 &R\\
02364-4751 \tablenotemark{b} & 02:38:13.1 & -47:38:11 &  0.098 & 12.10 
& 0,90 & 60,50 &B\\
04103$-$2838 & 04:12:19.5 & $-$28:30:24 &  0.117 & 12.55 & 89   & 60&R \\
04313$-$1649 & 04:33:37.1 & $-$16:43:32 &  0.268 & 12.55 & -1,89  & 120,120&R\\
05189$-$2524 & 05:21:01   & $-$25:21:46 &  0.043 & 12.09 & -1,89  & 200,160&R\\
06035-7102 \tablenotemark{b} & 06:02:54.0 & -71:03:10 &  0.0795 & 12.12 
& 65,153,153 & 60,50,60 &B\\
09039$+$0503 & 09:06:34.2 & $+$04:51:25 &  0.125 & 12.07 & -1,89  & 60,60&R\\
09111$-$1007 & 09:13:38.8 & $-$10:19:20 & 0.054  & 11.95 & 34,124 & 120,60&B \\
10190+1322 \tablenotemark{b} & 10:21:42   &  13:07:01 &  0.077 & 12.00 
& 64,149,149 & 40,40,40 &B\\
10565+2448 \tablenotemark{b} & 10:59:18.1 &  24:32:34 &  0.0431 & 12.02 
& -66,24 & 40,40 &B\\
11095-0238 \tablenotemark{b} & 11:12:03   & -02:54:18 &  0.106 & 12.20 
& 39,129 & 120,120 &B\\
11223$-$1244 & 11:24:50   & $-$13:01:13 &  0.199 & 12.59 & -1,89  & 80,80 &U\\
12071-0444 \tablenotemark{b} & 12:09:45.1 & -05:01:14 &  0.128  & 12.35 
& -1,89 & 60,60 &B\\
12112+0305 \tablenotemark{b} & 12:13:47   &  02:48:34 &  0.073 & 12.28 
& 37,99 & 60,60,40 &B\\
12540$+$5708 (Mrk 231) \tablenotemark{c} 
& 12:56:14.2 & $-$56:52:25 & 0.042 & 12.50 & 10,-30,-80 & 40,40,40 &R\\
13335-2612 \tablenotemark{b} & 13:36:22   & -26:27:31 &  0.125 & 12.06 
& -5 & 100 &B\\
13428$+$5608 (Mrk 273) \tablenotemark{c}
& 13:44:42.1 & $-$55:53:13 & 0.037 & 12.13 & 15,95 & 40,40&R\\
13451+1232 \tablenotemark{b} & 13:47:33   &  12:17:23 &  0.122 & 12.28 
& 104,13 & 80,120 &B\\
14070$+$0525 & 14:09:31.3 &  $+$05:11:31 & 0.264 & 12.76 & -1,89 & 120,120 &R\\
14348$-$1447 \tablenotemark{c}
& 14:37:38.3 & $-$15:00:23 & 0.0823 & 12.3 & 30 & 240 & B\\
14378$-$3651 \tablenotemark{c}
& 14:40:58.9 & $-$37:04:33 & 0.068 & 12.24 & -45 & 80 & R\\
15130$-$1958 & 15:15:55.2 & $-$20:09:17 &  0.109 & 12.09 & -171,-81 & 80,110 
&R\\
15250$+$3609 \tablenotemark{c} & 15:26:59.4 & $-$35:58:38 & 0.055 & 11.99 
& 45, -45 & 40,40 & R \\
15327$+$2340 (Arp 220) \tablenotemark{c} & 15:34:57.1 & 23:30:11 & 0.0181 & 
12.1 & 52, -91 & 120,120 & R \\
15462$-$0450 & 15:48:56.8 & $-$04:59:34 & 0.100 & 12.16 & 179,-91 & 180,160&R\\
16156+0146 \tablenotemark{b} & 16:18:08   &  01:39:21 &  0.132 & 12.04 
& -50,-51,40,40 & 60,60,60,60&B\\
16300+1558 \tablenotemark{b} & 16:32:20   &  15:51:49 &  0.242  & 12.63 
& -1,89 & 150,90&B\\
16504+0228 (NGC 6240)\tablenotemark{c} & 16:52:58.9 &  02:24:03 &  0.0245 & 
11.8 & -158,-31 & 20,20 &R\\
17208$-$0014 \tablenotemark{c}
& 17:23:21.9 & $-$00:17:00 & 0.0428 & 12.33 & 90,120 & 30, 30 &R\\ 
19254-7245 \tablenotemark{b} & 19:31:21.4 & -72:39:18 &  0.0617 & 12.00 
& -13,77 & 60,60 &B\\
20046-0623 \tablenotemark{b} & 20:07:19.3 & -06:14:26 &  0.0844 & 11.97 
& 69,159 & 60,60 &B\\
20087$-$0308 \tablenotemark{c}
& 20:11:23.2 & $-$02:59:54 & 0.106 & 12.40 & -45,45 & 40,40 & R\\
20414$-$1651 & 20:44:18.2 & $-$16:40:16 & 0.087  & 12.26 & 54,144 & 100,80&R \\
20551$-$4250 \tablenotemark{c}
& 20:58:26.9 & $-$42:39:06 & 0.0428 & 11.98 & -45,45 & 60,60 &R\\
21130-4446 \tablenotemark{b} & 21:16:18.5 & -44:33:38 &  0.0926 & 12.02 
& 33 & 40 &B\\
21208-0519 \tablenotemark{b} & 21:23:29 & -05:06:59 &  0.13   & 12.01 
& -164,109,109 & 60,60,60 &B \\
21219$-$1757 & 21:24:41.6 & $-$17:44:46 & 0.112  & 12.06 & -1,89  & 50,40 &R\\
21329-2346 \tablenotemark{b} & 21:35:45   & -23:32:36 &  0.125  & 12.09 
& 31 & 60 &B\\
21504$-$0628 & 21:53:05.5 & $-$06:14:50 &  0.078 & 11.92 & -39,59 & 60,60 &R\\
22491-1808 \tablenotemark{b} & 22:51:49.2 & -17:52:23 &  0.0778 & 12.09 
& -76,13,13 & 60,60,60 &B \\
23128-5919 \tablenotemark{b} & 23:15:46.8 & -59:03:15 &  0.045  & 11.96 
& -5,84,84 & 40,40,40 &B \\
23230$-$6926 & 23:26:03.6 & $-$69:10:19 & 0.106 
& 12.17 & -1,89  & 60,60 & R\\
23234+0946 \tablenotemark{b} & 23:25:56.2 &  10:02:50 &  0.128  & 12.05 
& -64,25 & 60,60 &B\\
23365$+$3604 \tablenotemark{c}
& 23:39:01.3 & $+$36:21:10 & 0.0645 & 12.09 & 45, -30 & 15, 40 &R\\
23578$-$5307 \tablenotemark{c}
& 00:00:23.6 & $-$52:50:28 & 0.125 & 12.10 & 107,14 & 60, 40 & R\\ 
\enddata

{\bf
\tablecomments{The coordinates, the redshift, the infrared luminosity, as 
well as the position angles and respective integration time for all our 
sources are presented in this Table. The data are presented in this paper 
unless otherwise noted.}

\tablenotetext{a}
{Pre- or post- coalescence classification, according to the scheme described
in \S~\ref{sec:selection}. ``B'' refers to a binary (pre-coalescence) source.
``R'' refers to a remnant (in post-coalescence). Remnants may also have two
nuclei, separated by no more than 1.5 kpc. ``U'' indicates that 
the classification is uncertain. }

\tablenotetext{b}
{Source presented in \pone.}

\tablenotetext{c}
{Source presented in Genzel et al. (2001) and Tacconi et al. (2002).}

}
\end{deluxetable}

\clearpage

\begin{deluxetable}{ccccccc}
\tablecolumns{7}
\tabletypesize{\footnotesize}
\tablewidth{0pt}
\tablecaption{\label{tab:structure} Derived structural and photometric 
parameters}
\tablehead{
\colhead{Galaxy} & \colhead{$R_{\rm eff}(H~band)$}  & \colhead{ellipticity}    
& \colhead{$\phi_{\alpha}$} & \colhead{$R_{\rm eff}(K~band)$}  
& \colhead{$\mu_{\rm gal}(H~band)$} & \colhead{$\mu_{\rm gal}(K~band)$} \\

\colhead{(IRAS)} & \colhead{(kpc)} & \colhead{} & \colhead{(\degree)}
& \colhead{(kpc)} & \colhead{mag $(\arcsec )^{-2}$} & 
\colhead{mag $(\arcsec )^{-2}$}
}   
\startdata
00091$-$0738 & 2.47($\pm 0.21$) & 0.220 &  15 & \nodata & \nodata & 16.8 
\tablenotemark{h}\\
00262$+$4251 \tablenotemark{a} & \nodata & \nodata & \nodata & 3.4 ($\pm 1.0$)
\tablenotemark{e} & \nodata & \nodata \\
00397$-$1312 \tablenotemark{b} & 2.04 ($\pm 0.76$) & 0.35 & -25 & \nodata
& 16.4  & 15.2  \tablenotemark{h} \\
F01004$-$2237 \tablenotemark{b} & 0.40 ($\pm 0.07$) & 0.02 & 26 & \nodata
& 14.3  & 12.9 \tablenotemark{k} \\ 
01388$-$4618 & 1.62 ($\pm 0.03$)& 0.074 & 55 & \nodata & \nodata & 14.7 
\tablenotemark{l,h} \\
01572$+$0009 & 1.31 ($\pm 0.10$) & 0.140 & 81 & 3.16 ($\pm 1.42$) 
\tablenotemark{c} & 14.6  \tablenotemark{d} & 13.7  \tablenotemark{k} \\ 
F02021$-$2103 \tablenotemark{b} & 5.38 ($\pm 3.95$) & 0.34 & 1 & \nodata
& 17.1  & 17.2  \tablenotemark{h} \\ 
04103$-$2838 \tablenotemark{b} & 1.61 ($\pm 0.12$) & 0.19 & -80 & \nodata
& 15.6  & 15.0  \tablenotemark{h} \\ 
04313$-$1649 \tablenotemark{b} & 4.04 ($\pm 0.89$) & 0.26 & 86 & 5.7 
\tablenotemark{f} & 18.5  & 18.0  \tablenotemark{h} \\ 
05189$-$2524 \tablenotemark{b} & 0.57 ($\pm 0.08$) & 0.06 & 77 & 0.79 
($\pm 0.02$) \tablenotemark{c} & 13.6  & 13.5  \tablenotemark{k}\\ 
09039$+$0503 \tablenotemark{b} & 1.62 ($\pm 0.90$) & 0.15 & -18 & \nodata
& 16.2  & 15.5  \tablenotemark{h} \\
11223$-$1244 & 3.83 ($\pm 0.38$) & 0.101 & 70 & \nodata & \nodata & 16.1 
\tablenotemark{h} \\ 
12540$+$5708 \tablenotemark{b} & 1.40 ($\pm 0.21$) & 0.08 & -11 & 0.2
\tablenotemark{e} & 13.4  & 13.2 \tablenotemark{k} \\ 
13428$+$5608 \tablenotemark{c}& 1.03 ($\pm 0.29$)& 0.498 & -77& 1.15($\pm0.24$)
\tablenotemark{c}& 14.5  \tablenotemark{d,h} & 14.1  \tablenotemark{d,h} \\
14070$+$0525 \tablenotemark{b} & 3.62 ($\pm 0.53$) & 0.17 & -19 & \nodata
& 17.2  & 16.2  \tablenotemark{h} \\
14378$-$3651 & 0.36 ($\pm 0.02$) & 0.031 & 20 & 0.67 \tablenotemark{e} & 
14.6 \tablenotemark{i} & 13.4  \tablenotemark{j,h} \\ 
15130$-$1958 \tablenotemark{b} & 1.62 ($\pm 0.27$) & 0.23 & -88 & \nodata
& 16.1  & 15.0  \tablenotemark{h} \\ 
15250$+$3609 \tablenotemark{c} & 2.10 ($\pm 0.09$)& 0.207& -72& 1.49($\pm0.12$)
\tablenotemark{c} & 16.9  \tablenotemark{d,h} & 16.5  \tablenotemark{d,h} \\ 
15462$-$0450 \tablenotemark{b} & 5.57 ($\pm 1.38$) & 0.048 & 62 & 1.6
 \tablenotemark{f} & 17.5  & 16.2  \tablenotemark{h} \\ 
17208$-$0014 & 1.69 ($\pm 0.08$)& 0.196 & 54& 1.63($\pm0.07$) \tablenotemark{c}
& 15.9  \tablenotemark{d,h} & 15.6  \tablenotemark{d,h}\\ 
20087$-$0308  & 1.87 ($\pm 0.22$)& 0.324 & -86 & \nodata & \nodata & 15.2 
\tablenotemark{j,h} \\ 
20414$-$1651 \tablenotemark{b} & 1.37 ($\pm 0.47$) & 0.67 & 3 & \nodata
& 16.2  & 15.7  \tablenotemark{h} \\ 
20551$-$4250 & 1.32 ($\pm 0.25$)& 0.113 & 84 & 3.3\tablenotemark{g} & 
\nodata & 16.9  \tablenotemark{j,h}\\ 
21219$-$1757 \tablenotemark{b} & 4.19 ($\pm 3.29$) & 0.14 & -50 & \nodata
& 16.9  & 15.6  \tablenotemark{h} \\  
21504$-$0628 & 1.95 ($\pm 0.26$) & 0.164 &  12 & \nodata & \nodata & \nodata\\ 
23230$-$6926 & 2.03 ($\pm 0.25$) & 0.177 &  41 & \nodata & 
16.9  \tablenotemark{i}  & 16.2 \tablenotemark{j,h} \\ 
23365$+$3604 \tablenotemark{a} & 4.8 ($\pm 1.0$) \tablenotemark{e} & \nodata 
& \nodata & \nodata & \nodata & \nodata \\
23578$-$5307 & 3.96 ($\pm 1.36$) & 0.447 & -80 & \nodata & \nodata & \nodata\\ 
Arp 220 \tablenotemark{d} & 3.00 & \nodata & \nodata & 0.69($\pm0.03$) 
\tablenotemark{c} & 17.4  \tablenotemark{d,h} & 16.9  \tablenotemark{d,h}\\
NGC 6240 \tablenotemark{d} & 1.44 & \nodata & \nodata & 0.92($\pm0.23$)
\tablenotemark{c} & 14.3  \tablenotemark{d,h} & 13.8  \tablenotemark{d,h} \\
00456$-$2904(SW) & 2.09 ($\pm 0.15$)& 0.067 & -88 & \nodata & 16.3  & 14.9\\
09111$-$1007(W) & 2.18 ($\pm 0.36$) & 0.397 & -26 & \nodata & \nodata & 
\nodata                   \\ 
14348-1447(NE) & 3.03 ($\pm 0.50$) & 0.237 & 4 & 3($\pm 1$)\tablenotemark{e} 
& 18.3  \tablenotemark{j} & 17.5  \tablenotemark{j}\\
14348-1447(SW) & 2.78 ($\pm 0.03$) & 0.113 & 7 & 3($\pm 1$)\tablenotemark{e} 
& 17.8 \tablenotemark{j} & 16.8  \tablenotemark{j}\\
\enddata

\tablecomments{
Structural and photometric properties of (mainly) ULIRG remnants. In this
table we present data for newly observed sources during our VLT large program
and we summarize observations carried out during our older normal programs
(\citealt{genzel01}; \citealt{tacconi02}) that are used in this work. The
sources in this Table together with those presented in \pone\ summarize the 
54 sources in our sample. The $H$-band structural parameters \reff, 
ellipticity, and $\phi_{\alpha}$ of the sources are derived from the ISAAC 
acquisition images unless otherwise noted in the first column. The angle 
$\phi_{\alpha}$ is tabulated here only for the first slit; for the second 
slit it can be computed using the angle between the two slit positions 
given in Table~\ref{tab:list}. The galaxy surface brightness within 
the effective radius, $\mu_{\rm gal}$ is derived from \cite{veilleux06} and 
\cite{kim02} respectively, unless otherwise noted. The $H$- and $K$-band 
$\mu_{\rm gal}$ data are presented after subtraction of the nuclear PSF for 
both bands; we indicate when individual PSF and galaxy magnitudes are not 
available in the literature. All data presented in this Table are derived 
from maximum-aperture photometry (enclosing the whole galaxy) and are not 
corrected for extinction effects.}

\tablenotetext{a}
{For the sources observed using NIRSPEC and the slit monitoring camera, the
structural parameters are not extracted; the slit projection drawn on the 
acquisition image does not allow for photometric analysis.}

\tablenotetext{b}
{Data taken from H-band NICMOS imaging by Veilleux et al. (2006). 
All quantities are PSF-subtracted and are converted to our cosmology.}

\tablenotetext{c}
{Structural parameters extracted from NICMOS imaging data, 
kindly provided to us by Nick Scoville.}

\tablenotetext{d}
{Based on Scoville et al. (2000).}

\tablenotetext{e}
{Effective radius from \cite{genzel01} or \cite{tacconi02}.}

\tablenotetext{f}
{Data from the 1 Jy sample analysis (\citealt{veilleux02}).}

\tablenotetext{g}
{Effective radius from \cite{rothberg04}.}

\tablenotetext{h}
{For these sources the relative strength of the PSF and the S\'ersic 
component fit to the underlying galaxy are not known. We assume that 
$L_{gal}(H~{\rm band})=0.64 L_{tot}(H~{\rm band})$ and that
$L_{gal}(K~{\rm band})=0.48 L_{tot}(K~{\rm band})$, (\citealt{colina01};
\citealt{surace99}; \citealt{veilleux06}).}

\tablenotetext{i}
{$H$-band magnitude taken from \cite{colina01}.}

\tablenotetext{j}
{magnitude taken from \cite{duc97}.}

\tablenotetext{k}
{$K$-band magnitude taken from \cite{surace99}.}

\tablenotetext{l}
{$K$-band magnitude taken from \cite{rigopoulou99}.}

\end{deluxetable}

\clearpage

\begin{deluxetable}{cccccccccc}
\tablecolumns{10}
\tabletypesize{\tiny}
\tablewidth{0pt}
\tablecaption{\label{tab:velocities} Stellar velocities and black hole masses}
\tablehead{
\colhead{Source} & \colhead{$\sigma$}   & 
\colhead{aper.\tablenotemark{a}} &
\colhead{aper.\tablenotemark{a}} &
\colhead{$V_{\rm rot}$(obs) \tablenotemark{b}}  &
\colhead{aper.\tablenotemark{c}} &
\colhead{$V_{\rm rot}{(\rm obs)} / \sigma$} &
\colhead{$M_{\rm BH}$  \tablenotemark{d}} & 
\colhead{$M_{\rm BH}$(Edd.)   \tablenotemark{e}} & 
\colhead{$\eta_{\rm Edd}$ \tablenotemark{f}} \\

\colhead{(IRAS)} & \colhead{(km s$^{-1}$)}  & \colhead{(\arcsec)}    
& \colhead{(kpc)} & \colhead{(km s$^{-1}$)} & \colhead{(\arcsec)}    
& \colhead{} & \colhead{(\msun)} & \colhead{(\msun)}
& \colhead{} 
}
\startdata
00091$-$0738     & 131   ($\pm$ 39)  & 0.45 & 1.1 & \nodata & \nodata   
& \nodata    & $2.46 \times 10^7$ &   $2.04 \times 10^7$ & 0.83 \\
00262$+$4251   & 170   ($\pm$ 15)  & \nodata & \nodata & $<15$ & \nodata & 
$<0.09$ & $7.02 \times 10^7$ &   $1.32 \times 10^7$ & 0.19 \\
00397$-$1312     & 106   ($\pm$ 26) & 0.36  & 2.0 & 49 ($\pm$ 17) & 0.58 & 
0.46  & $1.05 \times 10^7$ &   $1.05 \times 10^8$ & 9.94 \\
F01004$-$2237     & 132   ($\pm$ 29) & 0.29 & 0.7 & 22 ($\pm$ 13) & 0.60 
& 0.17      & $2.54 \times 10^7$ &   $2.29 \times 10^7$ & 0.90 \\
01388$-$4618 & 144   ($\pm$ 10)  & \nodata & \nodata & 130 ($\pm$ 15) & 
\nodata & 0.90      & $3.60 \times 10^7$ &   $1.32 \times 10^7$ & 0.37 \\
01572$+$0009  & 200   ($\pm$ 60)  & \nodata & \nodata & \nodata & \nodata 
& \nodata & $1.35 \times 10^8$ &   $4.46 \times 10^7$ & 0.33 \\
F02021$-$2103     & 143   ($\pm$ 21)  & 0.37 & 0.9 & 42 ($\pm$ 10) & 0.76 
& 0.29    & $3.50 \times 10^7$ &   $1.35 \times 10^7$ &  0.38 \\
04103$-$2838     & 129   ($\pm$ 40) & 0.29 & 0.7 & 4 ($\pm$ 5) & 0.31 
& 0.03         & $2.32 \times 10^7$ &   $1.86 \times 10^7$ &  0.80 \\
04313$-$1649     & 157   ($\pm$ 21) & 0.38  & 2.1 & 31 ($\pm$ 27) & 0.57 
&0.20      & $5.10 \times 10^7$ &   $4.67 \times 10^7$ &  0.92 \\
05189$-$2524     & 137   ($\pm$ 16) & 0.29 & 0.3 & 70 ($\pm$ 14) & 0.51 
& 0.51     & $2.95 \times 10^7$ &   $1.62 \times 10^7$ &  0.55 \\
09039$+$0503     & 183   ($\pm$ 38)  & 0.58 & 1.5 & \nodata & \nodata & 
\nodata      & $9.45 \times 10^7$ &   $1.55 \times 10^7$ &  0.16 \\
11223$-$1244     & 149   ($\pm$ 27)  & 0.73 & 3.0 & \nodata & \nodata & \nodata
& $4.13 \times 10^7$ &   $5.12 \times 10^7$ & 1.24    \\
12540$+$5708 & 120   ($\pm$ 10)  & \nodata & \nodata & 25 ($\pm10$) 
& \nodata & 0.21 & $1.73 \times 10^7$ &   $4.16 \times 10^7$ & 2.40 \\
13428$+$5608 & 285   ($\pm$ 30)  & \nodata & \nodata & 110 ($\pm20$) 
& \nodata & 0.39 & $5.61 \times 10^8$ &   $1.66 \times 10^7$ & 0.03 \\
14070$+$0525     & 139   ($\pm$ 21)  & 0.70 & 3.8 & 54 ($\pm$ 19) & 0.77  
& 0.39     & $3.13 \times 10^7$ &   $7.57 \times 10^7$ &  2.42 \\
14378$-$3651 & 153   ($\pm$ 10)  & \nodata & \nodata & 15 ($\pm10$) & 
\nodata & 0.10   & $4.60 \times 10^7$ &   $1.66 \times 10^7$ & 0.36 \\
15130$-$1958     & 177   ($\pm$ 39)  & 0.58 & 1.3 & 33 ($\pm$ 21) & 0.66 
& 0.19      & $8.26 \times 10^7$ &   $1.62 \times 10^7$ &  0.20 \\
15250$+$3609 & 150   ($\pm$ 10)  & \nodata & \nodata & 60 ($\pm15$) 
& \nodata & 0.40     & $4.25 \times 10^7$ &   $1.32 \times 10^7$ & 0.31 \\
15462$-$0450     & 169   ($\pm$ 38)  & 0.58 & 1.2 & \nodata & \nodata 
& \nodata      & $6.86 \times 10^7$ &   $1.90 \times 10^7$ & 0.28 \\
17208$-$0014  & 229   ($\pm$ 15) & \nodata & \nodata & 110 ($\pm$ 20) 
& \nodata & 0.48 & $2.33 \times 10^8$ &   $2.63 \times 10^7$ & 0.11 \\
20087$-$0308 & 219   ($\pm$ 14) & \nodata  & \nodata & 50 ($\pm$ 15) 
& \nodata & 0.23 & $1.94 \times 10^8$ &   $3.31 \times 10^7$ & 0.17 \\
20414$-$1651     & 187   ($\pm$ 32)  & 0.44 & 0.8 &  96 ($\pm$ 38) & 0.88 
& 0.51     & $1.03 \times 10^8$ &   $1.82 \times 10^7$ &  0.18 \\
20551$-$4250  & 140   ($\pm$ 15) & \nodata & \nodata & 40 ($\pm$ 10) & 
\nodata & 0.29   & $3.22 \times 10^7$ &   $1.32 \times 10^7$ & 0.41 \\
21219$-$1757     & 121   ($\pm$ 11) & 0.58 & 1.3 & \nodata & \nodata 
& \nodata      & $1.79 \times 10^7$ &   $1.51 \times 10^7$ &  0.84 \\
21504$-$0628     &  90   ($\pm$ 31) & 0.32  & 0.5 & 9 ($\pm$ 28) & 0.48 
& 0.10       & $5.45 \times 10^6$ &   $8.30 \times 10^6$ &  1.52 \\
23230$-$6926     & 143   ($\pm$ 14) & 0.48  & 1.1 & 23 ($\pm$ 13) & 0.73 
& 0.16      & $3.50 \times 10^7$ &   $1.44 \times 10^7$ &  0.41 \\
23365$+$3604  & 145   ($\pm$ 15)  & \nodata & \nodata & $<15$ & \nodata &
$<0.10$ & $3.71 \times 10^7$ &   $1.62 \times 10^7$ & 0.44 \\
23578$-$5307  & 190   ($\pm$ 70)  & \nodata & \nodata & \nodata & \nodata 
& \nodata      & $1.10 \times 10^8$ &   $1.66 \times 10^7$ & 0.15 \\
Arp 220 & 164 ($\pm$ 10) & \nodata\ & \nodata & 
185 ($\pm$ 30) \tablenotemark{g} & \nodata & 0.84 & $6.08 \times 10^7$ 
& $1.66 \times 10^7$ & 0.27\\
NGC 6240 & 229 ($\pm$ 43) \tablenotemark{h} & \nodata & \nodata 
& 240 ($\pm$ 108) \tablenotemark{h} & \nodata & 1.05 
& $2.33 \times 10^8$ & $8.30 \times 10^6$ & 0.04\\
00456$-$2904(SW) & 162 ($\pm$ 25) & \nodata & \nodata & 45 ($\pm$ 10) 
& \nodata& 0.28    & $5.79 \times 10^7$ &   $1.73 \times 10^7$ & 0.30 \\
09111$-$1007(W)  & 112   ($\pm$ 18)  & 0.29 & 0.3 & 68 ($\pm$ 16) & 1.17 
& 0.61      & $1.31 \times 10^7$ &   $9.53 \times 10^6$ &  0.73 \\
14348-1447(NE) & 170   ($\pm$ 14) & \nodata & \nodata & 60 ($\pm$ 20) & 
\nodata & 0.35   & $7.02 \times 10^7$ &   $1.92 \times 10^7$ &  0.27 \\
14348-1447(SW) & 150 ($\pm$ 25)  & \nodata & \nodata & 50 ($\pm$ 15) & 
\nodata & 0.33 & $4.25 \times 10^7$ &   $3.33 \times 10^7$ &  0.78 \\

\enddata

\tablecomments{ 
The stellar central velocity dispersion, rotational velocity, and the 
$V_{\rm rot}/\sigma$ ratio of the newly-observed sources are derived from 
the spectra of Fig.~\ref{fig:spectra} with the aid of 
the parameters of Table ~\ref{tab:structure}. The dynamical and Eddington 
black hole mass of each ULIRG and the ratio of the two are also presented 
here.}

\tablenotetext{a}
{Radial extent of the central aperture used for the extraction of $\sigma$,
tabulated both in angular and linear distances.} 

\tablenotetext{b}
{The observed rotational velocity value presented in this column is  
corrected for angular deviations from the major axis of rotation but not
for inclination effects.}

\tablenotetext{c}
{The center of the outer aperture used for the extraction of \vrot.
The annular width of this aperture is equal to the radius of the 
corresponding central aperture given in the third column.} 

\tablenotetext{d}
{Dynamical black hole masses estimated from their relation to the bulge 
dispersion (\citealt{tremaine02}).}

\tablenotetext{e}
{Eddington black hole mass, calculated by attributing 50\% of L$_{\rm IR}$ 
to the AGN.}

\tablenotetext{f}
{Ratio of Eddington black hole mass over dynamical black hole mass.}

\tablenotetext{g}
{We use as rotational velocity of Arp 220 that of the east component,
since that of the west component is only a lower limit (\citealt{genzel01}).}

\tablenotetext{h}
{Data are taken from \cite{tecza}. The average of the velocity dispersion
values at the two nuclei and the internuclear region is tabulated here.
This value is also close to the luminosity-weighted average of the two
nuclei which equals 225 \kms.}

\end{deluxetable}

\clearpage


\begin{figure*}
\centering
\includegraphics[height=20cm,width=16cm]{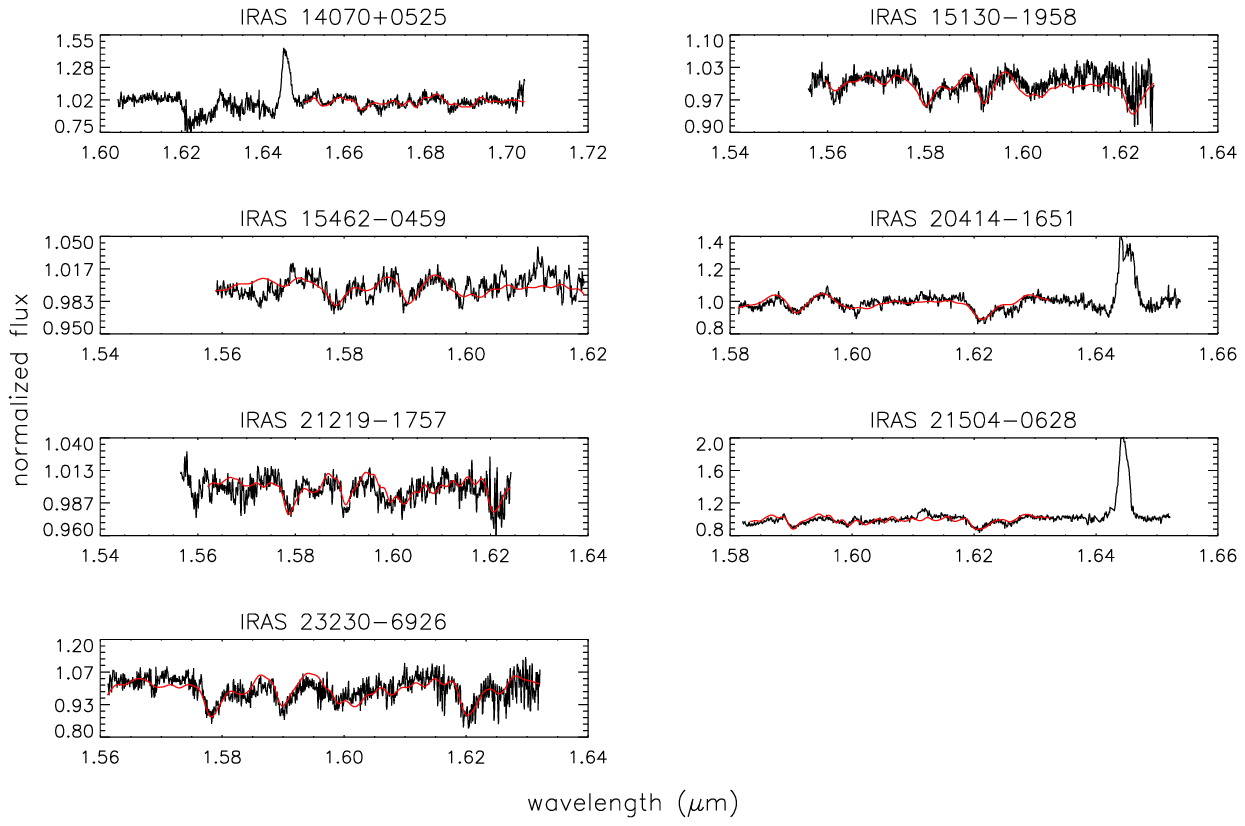}
Fig.~\ref{fig:spectra} continued.
\end{figure*}
\newpage

\end{document}